%% file: ANA-EXOT-2018-55-PAPER.tex
\documentclass[cernpreprint, atlasdraft=false,  UKenglish, texmf, orcidlogo]{atlasdoc}
\usepackage{atlaspackage}
\usepackage{multirow}
\usepackage{amsmath}
\usepackage[backref=false]{atlasbiblatex}

\usepackage[BSM=true,process=true]{atlasphysics}
\usepackage{adjustbox}
\graphicspath{{logos/}{figures/}}

\usepackage{ANA-EXOT-2018-55-PAPER-defs}

\AtlasTitle{Search for light neutral particles decaying promptly into collimated pairs of electrons or muons in $pp$ collisions at $\sqrt{s}$ = 13 \TeV with the ATLAS detector}

\AtlasAbstract{%
A search for a dark photon, a new light neutral particle, which decays promptly into collimated pairs of electrons or muons is presented.
The search targets dark photons resulting from the exotic decay of the Standard Model Higgs boson, assuming its production via the dominant gluon–gluon fusion mode.
The analysis is based on 140 fb$^{-1}$ of data collected with the ATLAS detector at the Large Hadron Collider from proton–proton collisions at a center-of-mass energy of 13~\TeV.
Events with collimated pairs of electrons or muons are analysed and background contributions are estimated using data-driven techniques.
No significant excess in the data above the Standard Model background is observed.
Upper limits are set at 95\% confidence level on the branching ratio of the Higgs boson decay into dark photons between 0.001\% and 5\%, depending on the assumed dark photon mass and signal model.
}

\AtlasRefCode{EXOT-2018-55}

\PreprintIdNumber{CERN-EP-2024-183}

\AtlasJournal{EPJC}
\AtlasJournalRef{\EPJC 85 (2025) 335}
\AtlasDOI{10.1140/epjc/s10052-025-13916-5}

\author{The ATLAS Collaboration}
\hypersetup{pdftitle={ATLAS document},pdfauthor={The ATLAS Collaboration}}

\begin{document}

\maketitle

%
%

%
\section{Introduction}
\label{sec:intro}
Hidden sectors near the weak scale that are motivated by naturalness, thermal dark matter, and electroweak baryogenesis are particularly compelling proposals for new phenomena beyond the Standard Model (SM)~\cite{ArkaniHamed:2008qp,Baumgart:2009tn,Falkowski:2014ffa,Curtin:2014cca}. Several minimal extensions of the SM introduce new symmetries and hidden sectors of particles that can be investigated at the Large Hadron Collider (LHC), assuming their interaction with SM particles through specific portals. Unstable dark states may be produced at colliders and decay into SM particles with sizeable branching ratios, depending on the dark sector's structure.

This paper focuses on a blueprint extension of the SM that considers an additional broken U$^\prime(1)$ gauge symmetry mediated by a massive vector boson, referred to as `dark photon' (\darky). The only interaction between the dark photon and SM particles is via kinetic mixing~\cite{Abel:2008ai} with the SM photon and the $Z$ boson, with a coupling denoted by $\epsilon$. Additionally, if a dark Higgs mechanism drives the spontaneous breaking of the new U$^\prime(1)$ gauge symmetry, the dark Higgs boson will generally have a coupling to the SM Higgs boson ($H$), leading to mixing between the two physical scalar states. The production of dark photons at the LHC can be enhanced by exotic decay modes of the Higgs boson, while its observation may be possible thanks to the kinetic mixing of the \darky with the SM photon.

The most stringent 95\% CL upper limit on the branching ratio for the SM Higgs boson decay into undetected final states is 12\%~\cite{HIGG-2021-23,CMS-HIG-22-001}, leaving ample room for the Higgs boson decay mode investigated in this paper.
In the absence of lighter hidden-sector states, a dark photon with a mass ($\mdarky$) up to a few \gev kinetically mixes with the SM photon and decays into leptons or light quarks.
Under this assumption, the dark photon decay branching ratios coincide with those of virtual SM photons, which are directly measured in $e^+e^-$ experiments~\cite{Zyla:2020zbs}.
The kinetic mixing parameter $\epsilon$ is related to the \darky mean proper lifetime. This paper targets values $\epsilon \gtrsim 10^{-5} - 10^{-3}$, which correspond to prompt \darky decays~\cite{Curtin:2014cca}.

The dark photon mass range targeted by this search goes from $\mathcal{O}(10\,\mev)$ to $\mathcal{O}(10\,\gev)$. Therefore, the small mass of the \darky relative to the Higgs boson mass implies that the dark photon decay products are highly collimated and can be identified as bundles of electrons or muons, referred to as \emph{Lepton-Jets} (\LJ) in the following. At masses greater than $\mathcal{O}(10\,\gev)$, the decay products of the dark photon become less collimated and can no longer be identified as \LJs.

This paper presents a search for \darky$\rightarrow e^+e^-$ and \darky$\rightarrow\mu^+\mu^-$ reconstructed as \LJs. The search is performed on proton--proton ($pp$) collision data at $\sqrt{s}=13\,\tev$, collected with the ATLAS detector at the LHC between 2015 and 2018 and corresponding to an integrated luminosity of $140\,\ifb$.
While sufficiently massive \darky can decay into quark-antiquark ($q\bar{q}$) or $\tau$-lepton pairs, these channels are not considered in this study, given the presence of neutrinos from $\tau$ decays and the overwhelming multi-jet background.
SM processes that can lead to a \LJ signature include the production of light vector mesons and off-shell photons. Due to the high cross-section of these background processes, the analysis is restricted to events where two \LJs are reconstructed.

Two minimal signal models where dark photons can be pair-produced are the Hidden Abelian Higgs Model (HAHM)~\cite{Curtin:2014cca} and the Falkowski-Ruderman-Volansky-Zupan (FRVZ) model~\cite{Falkowski:2010cm, Falkowski:2010gv}.
The former allows the Higgs boson decay into a pair of \darky, mainly via the diagram shown in Figure~\ref{fig:HAHM}, while the latter includes additional dark-sector fermions coupled to the \darky. The two dark-sector fermions ($f_d$) are produced by the Higgs boson decay, and they subsequently decay into a \darky and a Hidden Lightest Stable Particle (HLSP), as shown by the diagram in Figure~\ref{fig:FRVZ}. More complex scenarios where multiple pairs of \darky are produced are not considered in this paper, but the definition of the \LJ and the selection implemented in the analysis are designed to allow the interpretation of models with richer phenomenology.
\begin{figure}[tb]
\centering
\subfloat[]{\label{fig:HAHM}\includegraphics[height=4.5cm]{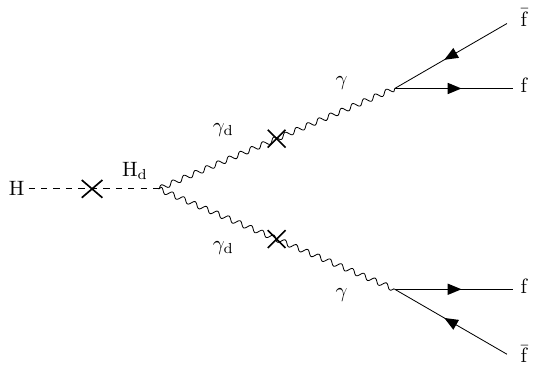}}
\subfloat[]{\label{fig:FRVZ}\includegraphics[height=4.5cm]{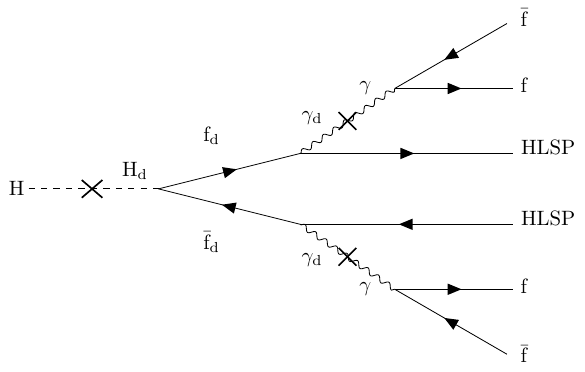}}
\caption{Representative diagrams for the Higgs boson decay in the (a) HAHM and (b) FRVZ models. In the HAHM model, dark photons (\darky) are produced from the decay of the Higgs boson, while in the FRVZ model, the Higgs boson decays in a pair of hidden fermions ($f_d$), both decaying into a dark photon and a stable hidden fermion (HLSP).}
\label{fig:feyn}
\end{figure}

Related searches targeting multiple production of \darky via the identification of \LJs were conducted by the ATLAS Collaboration on the Run~1 $pp$ collisions data at $\sqrt{s}=8\,\tev$~\cite{EXOT-2011-01,EXOT-2012-16,EXOT-2014-09}.
Two recent results by ATLAS~\cite{EXOT-2019-05,EXOT-2022-15} have investigated the production of long-lived dark photons using $\sqrt{s}=13\,\tev$ data, with masses in the $\mathcal{O}(10\,\mev)-\mathcal{O}(10\,\gev)$ range, probing the HAHM and FRVZ models and relying on the identification of \LJs from displaced \darky decays (referred to as `Dark-Photon-Jets' in the papers).
Other ATLAS searches targeted the scenario of dark photons with masses above $\mathcal{O}(1\,\gev)$ produced by Higgs boson decays, identifying resolved prompt~\cite{HDBS-2018-55} or displaced~\cite{EXOT-2017-03} decays of the \darky.
Similar searches were performed by the CMS Collaboration for prompt~\cite{CMS-EXO-11-013,CMS-EXO-12-012,CMS-HIG-13-010,CMS-HIG-19-007} and long-lived~\cite{CMS-EXO-20-014,CMS:2024qxz} dark photon production.
Other searches conducted by the LHCb~\cite{Aaij:2015tna,LHCb:2019vmc}, CMS~\cite{CMS-EXO-21-005} and FASER~\cite{FASER:2024-plb} Collaborations do not require the $H$-mediated production of dark photons, but assume kinetic mixing for both dark photon production and decay. These results target dark photon decays into muons and hence probe the mass regime $\mdarky > 2m_{\mu}$.
Several constraints on sub-\GeV dark photons are set, without assuming the production from Higgs boson decays, by beam-dump and fixed-target experiments~\cite{Bross:1989mp,Reece:2009un,Bjorken:2009mm,Essig:2010gu,Blumlein:2011mv,Merkel:2011ze,Abrahamyan:2011gv,Gninenko:2012eq,Adlarson:2013eza,Blumlein:2013cua,Agakishiev:2013fwl},
by measurements of the anomalous magnetic moment of electrons and muons~\cite{Pospelov:2008zw,Davoudiasl:2012ig,Endo:2012hp} and by astrophysical observations~\cite{Dreiner:2013mua,Chang:2018rso}.
Experiments at $e^+e^-$ colliders~\cite{Aubert:2009cp,Archilli:2011zc,BABAR:PRL2012,Babusci:2012cr,Lees:2014xha,Lees:2015rxq,TheBelle:2015mwa,KLOE2:2015-plb,Won:2016pjz,Ablikim:2017aab,Ablikim:2018bhf,Belle:2020,Belle2:PRL2023} also constrain the production of dark photons with masses smaller than $\mathcal{O}(10\,\GeV)$.

This paper significantly improves the previous Run~1 result~\cite{EXOT-2014-09} thanks to a novel event selection, optimised considering multiple \darky mass scenarios, to a new background estimate based on a bump-hunt strategy, and to the larger integrated luminosity and increased Higgs boson production cross section at $\sqrt{s}=13\,\tev$.
The HAHM model, not considered by the previous search, is also constrained for the first time using prompt \LJs, probing a complementary $\mdarky$ range with respect to that probed by Ref.~\cite{HDBS-2018-55}.


\FloatBarrier
%

%

\section{ATLAS detector}
\label{sec:detector}
\newcommand{\AtlasCoordFootnote}{%
ATLAS uses a right-handed coordinate system with its origin at the nominal interaction point (IP)
in the centre of the detector and the \(z\)-axis along the beam pipe.
The \(x\)-axis points from the IP to the centre of the LHC ring,
and the \(y\)-axis points upwards.
Polar coordinates \((r,\phi)\) are used in the transverse plane,
\(\phi\) being the azimuthal angle around the \(z\)-axis.
The pseudorapidity is defined in terms of the polar angle \(\theta\) as \(\eta = -\ln \tan(\theta/2)\) and is equal to the rapidity
$ y = \frac{1}{2} \ln \left( \frac{E + p_z c}{E - p_z c} \right) $ in the relativistic limit.
Angular distance is measured in units of \(\Delta R \equiv \sqrt{(\Delta y)^{2} + (\Delta\phi)^{2}}\).}

The ATLAS detector~\cite{PERF-2007-01} at the LHC covers nearly the entire solid angle around the collision point.\footnote{\AtlasCoordFootnote}
It consists of an inner tracking detector surrounded by a thin superconducting solenoid, electromagnetic and hadronic calorimeters,
and a muon spectrometer incorporating three large superconducting air-core toroidal magnets with eight coils each.

The inner-detector system (ID) is immersed in a \qty{2}{\tesla} axial magnetic field
and provides charged-particle tracking in the range \(|\eta| < 2.5\).
The high-granularity silicon pixel detector covers the vertex region and typically provides four measurements per track,
the first hit generally being in the insertable B-layer (IBL) installed before Run~2~\cite{ATLAS-TDR-19,PIX-2018-001}.
It is followed by the SemiConductor Tracker (SCT), which usually provides eight measurements per track.
These silicon detectors are complemented by the transition radiation tracker (TRT),
which enables radially extended track reconstruction up to \(|\eta| = 2.0\).
The TRT also provides electron identification information
based on the fraction of hits (typically 30 in total) above a higher energy-deposit threshold corresponding to transition radiation.

The calorimeter system covers the pseudorapidity range \(|\eta| < 4.9\).
Within the region \(|\eta|< 3.2\), electromagnetic calorimetry is provided by barrel and
endcap high-granularity lead/liquid-argon (LAr) calorimeters,
with an additional thin LAr presampler covering \(|\eta| < 1.8\)
to correct for energy loss in material upstream of the calorimeters.
Hadronic calorimetry is provided by the steel/scintillator-tile calorimeter,
segmented into three barrel structures within \(|\eta| < 1.7\), and two copper/LAr hadronic endcap calorimeters.
The solid angle coverage is completed with forward copper/LAr and tungsten/LAr calorimeter modules
optimised for electromagnetic (EM) and hadronic energy measurements respectively.

The muon spectrometer (MS) comprises separate trigger and
high-precision tracking chambers measuring the deflection of muons in a magnetic field generated by the superconducting air-core toroidal magnets.
The field integral of the toroids ranges between \num{2.0} and \qty{6.0}{\tesla\metre}
across most of the detector.
Three layers of precision chambers, each consisting of layers of monitored drift tubes, cover the region \(|\eta| < 2.7\),
complemented by cathode-strip chambers in the forward region, where the background is highest.
The muon trigger system covers the range \(|\eta| < 2.4\) with resistive-plate chambers in the barrel, and thin-gap chambers in the endcap regions.

The luminosity is measured mainly by the LUCID--2~\cite{LUCID2} detector that records Cherenkov light produced in the quartz windows of photomultipliers located close to the beampipe.

Events are selected by the first-level trigger system implemented in custom hardware,
followed by selections made by algorithms implemented in software in the high-level trigger~\cite{TRIG-2016-01}.
The first-level trigger accepts events from the \qty{40}{\MHz} bunch crossings at a rate below \qty{100}{\kHz},
which the high-level trigger further reduces in order to record complete events to disk at about \qty{1}{\kHz}.

A software suite~\cite{SOFT-2022-02} is used in data simulation, in the reconstruction
and analysis of real and simulated data, in detector operations, and in the trigger and data acquisition
systems of the experiment.


\FloatBarrier
%

%

%
\section{Data and simulated event samples}
\label{sec:samples}
Data are collected by the ATLAS detector during Run~2 of the LHC (2015--2018) at $\sqrt{s}=13\,\tev$ and correspond to an integrated luminosity of $140\,\ifb$. The highest peak instantaneous luminosity reached $2.1\times 10^{34}$~cm$^{-2}$s$^{-1}$, with an average number of inelastic interactions per bunch crossing ranging from 13.4 to 37.8, depending on the data-taking year.
These are collected using a set of triggers that require the presence of electrons~\cite{TRIG-2018-05} or muons~\cite{TRIG-2018-01} with thresholds in the transverse energy for electrons and the transverse momentum ($\pt$) for muons that are in the range of 6--26~\GeV, depending on the lepton flavour, the number of leptons and the data-taking period~\cite{TRIG-2019-03}.
Stringent data quality requirements~\cite{DAPR-2018-01} are applied, ensuring the optimal operation of all sub-detectors and stable-collision mode of the LHC beams.

Simulated Monte Carlo (MC) event samples are used to optimise the analysis selections and characterise the signal and backgrounds.
The signal samples consist of events where dark photons are produced, according to the HAHM and FRVZ models, via the decay of the SM Higgs boson with mass set to $125\,\gev$ and considering only the dominant gluon--gluon fusion production mechanism.
Several samples with different \darky masses were generated, varying from $17\,\mev$ to $20\,\gev$. For the HAHM model, dark photon masses were set to 17, 100, 400, 2000, 10000 and 20000\,\mev, while for the FRVZ model the generated samples include dark photon masses of 17, 30, 60, 100, 240, 400, 900, 2000, 6000, 10000\,\mev.
The decays of the Higgs boson into dark photons through dark fermions or directly into two dark photons were simulated at the matrix-element level during the generation.
The branching ratios of the dark photon decay into leptons were enhanced during event generation and the events are weighted in order to match the expected decay branching ratios of a virtual SM photon.
In the FRVZ model, the mass of \fd~was chosen to be small relative to the Higgs boson mass, and far from the kinematic threshold at $m_{f_{\mathrm{d}}} = m_{\mathrm{HLSP}} + m_{\gamma_{\mathrm{d}}}$. The values of the dark fermions masses have a negligible impact on the analysis results. In the HAHM, the Higgs boson can decay directly into a pair of dark photons leading to more boosted final states.
Simulated signal events were generated with \MGNLO[2.2.3]~\cite{Alwall:2014hca} interfaced to \Pythia[8.186]~\cite{Sjostrand:2007gs} for the parton showering and hadronisation. The matrix-element calculation was performed at tree level and the parton distribution function (PDF) set used for the generation was~\NNPDF[2.3lo]~\cite{Ball:2012cx}.
Signal samples were normalised to a total cross-section of 48.61~pb~\cite{Cepeda:2019klc,Anastasiou:2016cez}, using dedicated cross-sections calculations at NNLO in QCD and including electroweak corrections at NLO.

Simulated SM background samples were generated in order to optimise the event selection and include $Z$ boson production in association with jets ($Z$+jets) and top-quark pair production (\ttbar).
The production of $Z$+jets was simulated  with the \SHERPA[2.2.1]~\cite{Bothmann:2019yzt} generator using NLO matrix elements for up to two partons, and leading-order (LO) matrix elements for up to four partons calculated with the \textsc{Comix}~\cite{Gleisberg:2008fv} and \OPENLOOPS~\cite{Buccioni:2019sur,Cascioli:2011va,Denner:2016kdg} libraries. They were matched with the \SHERPA parton shower~\cite{Schumann:2007mg} using the \MEPSatNLO prescription~\cite{Hoeche:2011fd,Hoeche:2012yf,Catani:2001cc,Hoeche:2009rj} using the set of tuned parameters developed by the \SHERPA authors. The \NNPDF[3.0nnlo] set of PDFs~\cite{Ball:2014uwa} was used and the samples were normalised to a NNLO prediction~\cite{Anastasiou:2003ds}.
The production of \ttbar events was modelled using the \POWHEGBOX[v2]~\cite{Frixione:2007nw,Nason:2004rx,Frixione:2007vw,Alioli:2010xd} generator at NLO with the \NNPDF[3.0nlo] PDF set~\cite{Ball:2014uwa} and the \hdamp parameter set to 1.5 times the mass of the top quark~\cite{ATL-PHYS-PUB-2016-020}.  The events were interfaced to \PYTHIA[8.230]~\cite{Sjostrand:2014zea} to model the parton shower, hadronisation, and underlying event, with parameters set according to the A14 tune~\cite{ATL-PHYS-PUB-2014-021} and using the \NNPDF[2.3lo] PDF set. The decays of bottom and charm hadrons were performed by \EVTGEN[1.6.0]~\cite{Lange:2001uf}.

All MC simulated events were processed through a full simulation of the ATLAS detector geometry and detector response~\cite{SOFT-2010-01} using the \GEANT~\cite{Agostinelli:2002hh} toolkit. This simulation accounts for multiple $pp$ interactions per bunch crossing (pile-up), as well as the detector response to interactions in bunch crossings before and after the one producing the hard interaction.
The multiple $pp$ interactions were included using simulated events generated with \PYTHIA[8.186]~\cite{Sjostrand:2007gs} using the \NNPDF[2.3lo] PDF set and the A3 minimum-bias tune~\cite{ATL-PHYS-PUB-2016-017}. Simulated events are weighted to reproduce the distribution of the average number of interactions per bunch crossing observed in data.


\FloatBarrier
%

%

%
\section{Event reconstruction}
\label{sec:reco}
The presence of at least one collision vertex, reconstructed from at least two ID tracks with $\pt>500\,\mev$, is required for each event~\cite{ATL-PHYS-PUB-2015-026}. When multiple vertices satisfy this requirement, the one with the largest $\sum\pt^2$ is selected as the primary vertex of the event.

Electrons are identified by associating a cluster of energy deposits in the electromagnetic calorimeter to at least one track in the ID.
Their \pt must be greater than $4.5\,\gev$ and they must be found within $\abseta<2.47$, with the exclusion of the transition region between the barrel and endcap electromagnetic calorimeter, defined by $1.37 < \abseta < 1.52$. Electrons must also satisfy the \emph{Medium} identification working point defined in Ref.~\cite{EGAM-2018-01}.
In addition, for tracks associated to electrons the significance on the transverse impact parameter is required to be $|d_0|\ /\ \sigma_{d_0} < 5$ and the longitudinal impact parameter must be $|\Delta z_0\sin{\theta}| < 0.5\,\mm$, where both impact parameters are computed with respect to the primary vertex.
Track-based and calorimeter-based isolation are required according to the \emph{Loose} criterion defined in Ref.~\cite{EGAM-2018-01}.
Electromagnetic showers from light \darky decaying in two electrons are often merged into a single cluster, which is then reconstructed as a single electron with two associated tracks.
The impact of the isolation requirement, when two close-by electrons are identified as a single EM shower, is found to be negligible for electrons identified from simulated \darky decays into electrons.

Muon tracks in the ID and the MS are used in a \emph{Combined} muon fit~\cite{MUON-2018-03}. For each muon, a minimum \pt of $3\,\gev$ is required as well as $\abseta < 2.5$ and the \emph{Loose} identification working point~\cite{MUON-2018-03} must be satisfied. Muon ID tracks are required to satisfy $|d_0|\ /\ \sigma_{d_0} < 3$ and $|\Delta z_0\sin{\theta}| < 0.5\,\mm$.
Muon isolation requirements are based on the presence of track-based or calorimeter-based energy contributions around a muon.
Muons used for the definition of the regions where the background modelling is performed are required to satisfy the \emph{PFlowLoose} isolation requirement with variable cone radius~\cite{MUON-2018-03}.
Such requirements are found to be inefficient when applied on muons originating from boosted \darky$\rightarrow\mu\mu$ decays, due to the contribution of one muon to the isolation cone of the other. A custom isolation variable, defined without considering the contribution due to nearby muons found in a $0.4$ cone, is applied to muons used in the \LJ reconstruction, leading to an increase of up to 80\% in the selection efficiency with respect to standard requirements.

Hadronic jets are reconstructed from clusters of energy deposits in the calorimeter~\cite{ATL-PHYS-PUB-2015-036} using the anti-$k_t$ algorithm~\cite{Cacciari:2008gp,Fastjet}, with a radius parameter $R=0.4$.
The energy calibration procedure described in Ref.~\cite{PERF-2016-04} is also applied and the jets are required to have  $\pt>20\,\gev$, $\abseta < 2.5$ and they must satisfy the \emph{Loose} selection defined in~\cite{ATLAS-CONF-2015-029}.
To suppress jets from pile-up interactions, jets with $\pt<60\,\gev$, $\abseta < 2.5$ are required to satisfy a selection on a multivariate jet vertex tagger~\cite{PERF-2014-03} based on tracking information.

In order to avoid double-counting of objects, electrons sharing an ID track with muons are removed. Reconstructed jets are discarded if they are found within a $\Delta R=0.2$ cone around a lepton (and only if they have less than three ID tracks in case of muons). Remaining jets are retained against leptons if they satisfy $\Delta R < \textrm{min}(0.4, 0.04 + 10\,\gev/\pt^{\ell})$, where $\pt^{\ell}$ is the \pt of the lepton.

\LJs are identified from collimated muons or electrons originating from the decay of light \darky. Two exclusive types of \LJs are identified: \emph{muon~\LJs} (\mLJs) or \emph{electron~LJs} (\eLJs). Cambridge--Aachen clustering~\cite{Dokshitzer:1997in} is adopted, starting from the first lepton in the collection ordered by \pt and adding the same-flavour ones found within a $\Delta R=0.4$ cone around the initial one. For each lepton added to the \LJ, the momentum axis is recomputed as the vector sum of the four-momentum of each constituent. The procedure is repeated until no other particle can be added and until no other \LJ can be reconstructed. If a different-flavour lepton is found within the cone of a given \LJ, such \LJ is discarded, in order to ensure orthogonality between the two types. The \LJ definition is inclusive in the number of leptons to allow the interpretation of the analysis in the context of more complex models, as mentioned in Section~\ref{sec:intro}.
\LJs are expected to be reconstructed with all the electrons or muons originating from the decay of a neutral particle, hence the sum of the charges of the \LJ components (muons in case of \mLJs and the associated ID tracks in case of \eLJs) is required to be zero.

Muon \LJs contain at least two muons and target the signature where one or more \darky decay into muons.
Electron \LJs are required to contain at least one electron and at least two ID tracks associated to electrons, to accommodate for the scenarios in which the electrons from the \darky decay are identified as two resolved objects, as well as the one in which the EM showers from the two electrons are merged and identified as a single electron with two associated tracks.
In addition, only \eLJs with $|\eta|< 1.37$, for which the leading track has a \pt >5 \gev, are retained for the selection.

The mass of a \LJ is defined as the invariant mass of its components:
for \mLJs this corresponds to the invariant mass of the muon constituents while, for \eLJs, the definition depends on the number of electron objects that are clustered.
For an \eLJ reconstructed from two or more electrons, its mass is taken as the invariant mass of the system of electrons.
On the other hand, for an \eLJ containing two ID tracks associated to one electron, the invariant mass is computed by using the associated ID tracks. This definition was found to be optimal in order to correctly reconstruct the mass of the \darky in simulated signal events.
Since this search targets the pair-production of two particles decaying into electron/muon pairs, it is also expected that the masses of the two \LJs that are reconstructed are similar, hence the absolute value of the ratio between the difference and the sum of the masses of the \LJs (hereafter referred to as `mass imbalance'), is used when defining the analysis regions.

Two additional variables are defined exclusively for \eLJs.
The imbalance in \pt (\ptimb) is defined as the absolute value of the ratio between the difference and the sum of the \pt of the ID tracks associated to an \eLJ. If more than two tracks are associated to an \eLJ, the \ptimb is computed by considering the two leading tracks in \pt with opposite charge.
The \ptimb is expected to be smaller for \eLJ originating from resonance decays, with respect to \eLJs formed from prompt electrons matched to additional ID tracks.
On the other hand, the shape of a merged EM shower originating from two close-by electrons is expected to be wider in $\phi$ for an \eLJ originating from a \darky decay, compared to that of a prompt electron matched to an additional track.
Hence, the ratio between the energy of the EM shower, in the cells around the most energetic energy cluster measured in a $(\eta \times \phi)=3\times 3$ region and the one measured $3\times7$ region (hereafter referred to as \rphi) is used when defining the signal region (SR).

The presence of at least two \LJs is required in each event and the event selection is performed on the leading \LJ in \pt and the \LJ with largest \delphi from the leading \LJ (denoted  farthest \LJ).
The event selection continues with two analysis channels, which are defined depending on the type of the leading and farthest \LJs. The \emph{muon channel} includes events where at least one \mLJ is found and is described in Section~\ref{sec:muLJ_analysis}, while the \emph{electron channel} includes events where both \LJs are of the electron type and is described in Section~\ref{sec:eLJ_analysis}. Events with a single reconstructed \LJ are not considered as the constraints they provide are not competitive due to the large background yields.


\FloatBarrier
%

%

%
\section{Event selection and background estimation in the muon channel}
\label{sec:muLJ_analysis}
Events with at least one reconstructed \mLJ are required to satisfy single-lepton or multi-lepton (dilepton and trilepton) triggers, including triggers based on mixed lepton flavour.
Offline leptons used in the \LJ reconstruction are required to be found within a \delr$=0.2$ cone around the trigger seed and, in the case of multi-lepton triggers, the leptons matching the online objects must belong to two different \LJs.
Moreover, a selection on the \pt, identification and isolation of the leptons matching the trigger is applied, tighter than the requirements applied online, to be on the trigger efficiency plateau~\cite{TRIG-2018-05,TRIG-2018-01}.
In order to avoid overlaps between events selected with the di-muon and the tri-muon triggers, the latter ones are utilised only for events outside the fiducial region of the di-muon triggers.

Two orthogonal regions are defined, based on the \LJ multiplicity: events with at least two reconstructed \mLJs and no \eLJs, the \twomLJ region, and events where a \mLJ and an \eLJ are reconstructed, the \mixLJ region.
While no further requirement is applied in the \twomLJ region, the \mixLJ region requires \ptimb$<0.8$ for \eLJ and that the two \LJs are back-to-back, with a minimum separation in the azimuthal angle of at least two radians, in order to improve the purity of the selection. The requirements are summarised in Table.~\ref{tab:muonicSR}.

\begin{table}[!tb]
\caption{Definition of the muon signal regions. In signal regions requiring events with at least two \LJs, only the leading \LJ and the farthest \LJ are considered for the event classification.}
\def\arraystretch{1.2}
\begin{center}
\begin{tabular}{l r r}
\noalign{\smallskip}\hline\noalign{\smallskip}
Requirement / Region               & \twomLJ        & \mixLJ    \\
\noalign{\smallskip}\hline\noalign{\smallskip}
Number of \mLJs                    & $\ge 2$           & $\ge 1$         \\
Number of \eLJs                    & 0                & $\ge 1$         \\
\hline
muon triggers 		       & yes            & yes       \\
electron-muon triggers             & --              & yes       \\
electron triggers                  & --              & yes       \\
\hline
\eLJ \ptimb                        & --              & $<0.8$    \\
$\Delta\phi(\mu \text{LJ},e\text{LJ})$           & --              & $>2$      \\
\hline
\end{tabular}
\label{tab:muonicSR}
\end{center}
\end{table}

The final result of the search in the \mLJs regions is obtained by fitting the invariant mass of the \mLJ.
In the \twomLJ region, there are two entries per event, corresponding to the invariant masses of both \mLJs.
In the \mixLJ region, only the \mLJ invariant mass is utilised, as the overlapping electron showers in the calorimeter prevented a well-defined \eLJ mass reconstruction.
The SM background affecting these search regions is characterised by a non-resonant component due to the production of muon pairs from virtual photons decays, plus a resonant component due to the pair-production of low-mass resonances, such as the \jpsi meson, decaying into muons.
For the signal, the \mLJ mass distribution is modelled using a parametric function, with parameters interpolated to include signal masses for which MC simulation is not available. A double-sided Crystal Ball function~\cite{Oreglia:1980cs} is found to be an excellent choice for modelling the shape of the signal invariant mass distribution.

Examples of simulated distributions of the number of \mLJ candidates within the mass range used for this region are shown in Figure~\ref{fig:mLJmass}. The \mLJ invariant mass resolution ranges from 10 to $300\,\mev$ in the investigated mass range.

\begin{figure}[tb!]
\centering
\includegraphics[width=.49\textwidth]{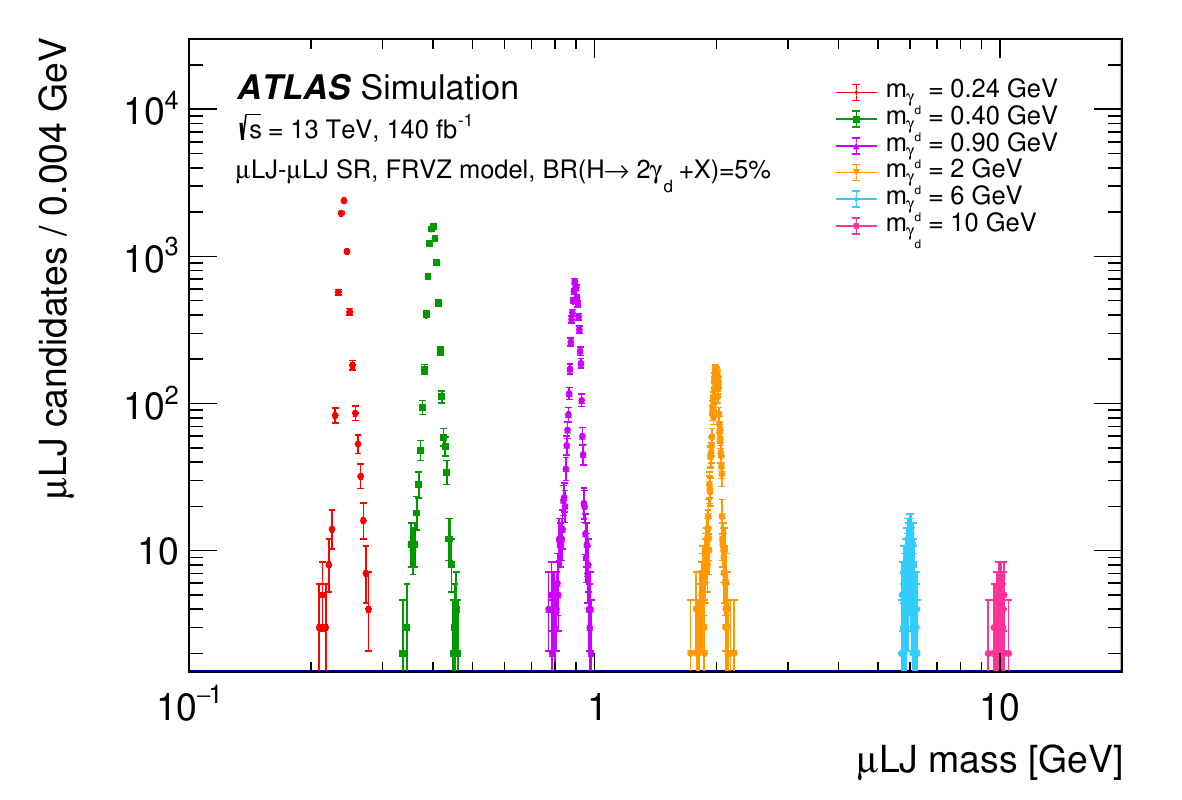}
\caption{Simulated distributions of the number of \mLJ candidates for a selection of \darky mass values. The shape and normalisation of the distributions are extracted from the parameterisation obtained for \mLJ-\mLJ SR, using the FRVZ model and assuming a branching ratio of the Higgs boson decay to dark photons of 5\%.}
\label{fig:mLJmass}
\end{figure}

The background distribution corresponds to a smooth falling spectrum and vector meson resonances.
This enables the background estimate to be performed using a data-driven method, in which the continuum shape is parameterised by an analytical function derived from two control regions (CRs).
The CRs are obtained by selecting events with one reconstructed \mLJ, no reconstructed \eLJs, and two additional pairs of muons or electrons separated by $\Delta R \geq 1.8$, which are not used for the reconstruction of \LJs. The \mLJ distribution is then used to characterise the modelling of the \mLJ distribution in the SR.
The CR built from one \mLJ and two muons is used to characterise the modelling of the \twomLJ search region, while for the \mixLJ region the CR built from one \mLJ and two additional electrons is used.
These additional leptons must fulfil the same requirements as the \mLJ or \eLJ constituents, ensuring that the CR closely resembles the SR.
In order to minimise any signal contamination in the CRs, a requirement on the mass imbalance of the \mLJ and the di-muon system, or the \mLJ and di-electron system is applied, requiring it to be greater than 0.2, or 0.6, respectively.

The parametric form of the \mLJ distribution for the background has two components: a double exponential function to describe the bulk non-resonant distributions, and Gaussian probability functions to capture the  $\phi(1020)$, $\jpsi$, and $\psi(2S)$ resonances. It is parameterised as follows:
\begin{equation}
\label{eq:bkg_pdf}
\begin{aligned}
B(m_{\mu \text{LJ}}) =& N_{\mathrm{exp1}}e^{-m_{\mu \text{LJ}}/\tau_2}+N_{\mathrm{exp2}}e^{-m_{\mu \text{LJ}}/\tau_1} \\
&    + N_{\jpsi}e^{-\left(\frac{ m_{\mu \text{LJ}}- \mu_{\jpsi}}{\sigma_{\jpsi}}\right)^2}+N_{\psi(2S)}e^{-\left(\frac{ m_{\mu \text{LJ}}- \mu_{\psi(2S)}}{\sigma_{\psi(2S)}}\right)^2}+N_{\phi}e^{-\left(\frac{ m_{\mu \text{LJ}}- \mu_{\phi}}{\sigma_{\phi}}\right)^2} ,
\end{aligned}
\end{equation}
where $N_i$, with $i = \mathrm{exp1},\mathrm{exp2},\phi,\jpsi,\psi(2S)$, are the normalisation factors associated to the $i$-th SM background process relative to the total amount of events; $\tau_1 $ and $\tau _2$ are the parameters of the two exponential distributions.
The parameters $\mu_{\text{res}}$ with $\text{res}=\phi,\jpsi,\psi(2S)$ are fixed and set to $1.02$, $3.097$ and $3.69\,\gev$, respectively~\cite{Zyla:2020zbs}. The parameters $\sigma _\text{res}$, with $\text{res} = \phi,\jpsi,\psi(2S)$ are fixed to the fitted values in the CR.
This functional form is chosen for its ability to accurately model the background shape in several control regions.

The bias from the functional form choice is evaluated by fitting signal and background contributions to data templates in signal-free regions. This \emph{spurious signal} estimate helps gauge biases inherent in the method~\cite{HIGG-2019-23}. Various templates are created using data or simulation in the control region, and used as probability distribution functions to generate expected distributions in the SR. The extracted values of the signal yields for each \darky mass assumption define an envelope approximating the method's bias, which is included as a systematic uncertainty in the final fit.

Post-fit \mLJ mass distributions in the \mLJ--\mLJ and \mixLJ signal regions are shown in Figure~\ref{fig:mLJpostfit}.
In this figure, a representative \darky signal with a mass of $1\,\gev$ is shown, assuming a branching ratio of \HtwogdHAHM of 0.5\%.

\begin{figure}[tb!]
\centering
\subfloat[]{\includegraphics[width=.49\textwidth]{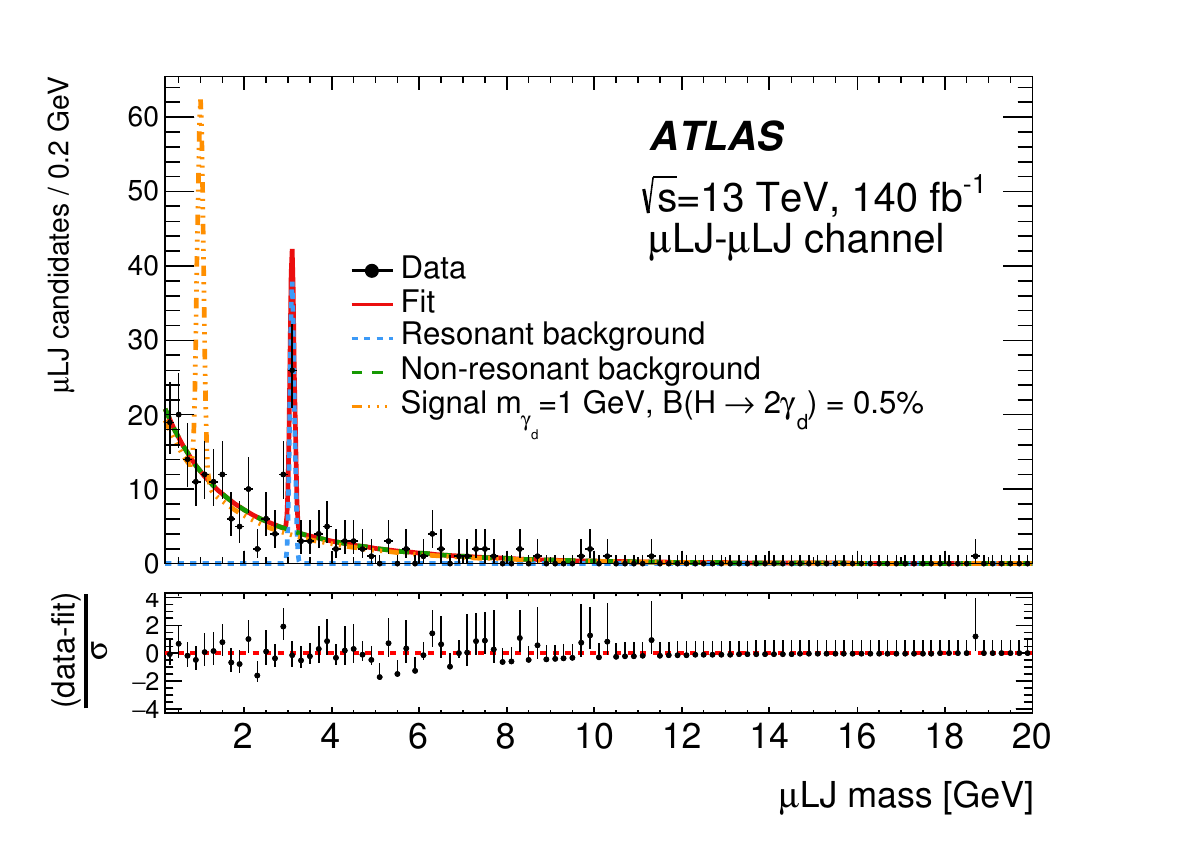}}
\subfloat[]{\includegraphics[width=.49\textwidth]{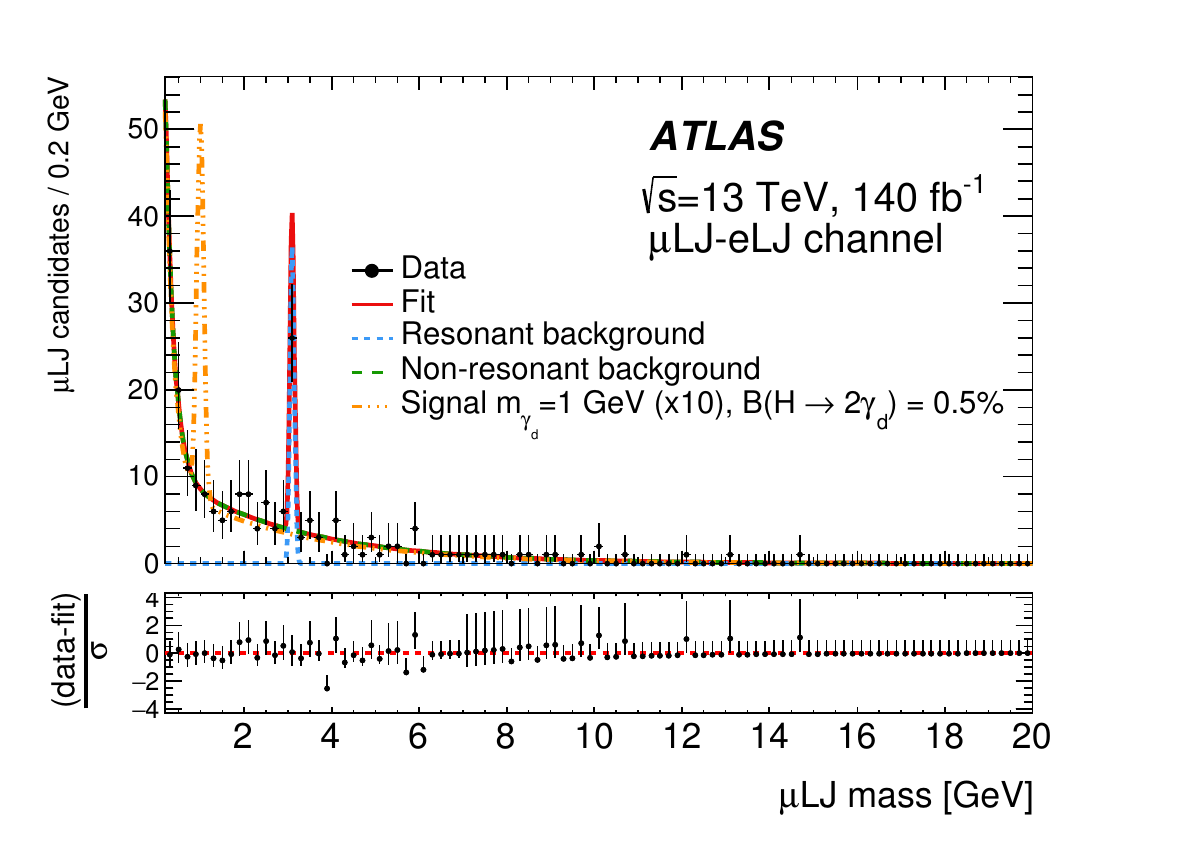}}
\caption{The background-only fit (solid red line) with its background (dashed blue and green lines) and signal (dot-dashed orange line) components of the \mLJ mass distributions for the (a) \mLJ--\mLJ and (b) \eLJ--\mLJ regions. For the \mLJ--\mLJ region, both the \mLJs are included. A signal distribution for a dark photon mass of 1~\gev is overlaid, assuming the HAHM model and a branching ratio of the Higgs boson to dark photons of 0.5\%.
}
\label{fig:mLJpostfit}
\end{figure}


\FloatBarrier
%

%

%
\section{Event selection and background estimation in the electron channel}
\label{sec:eLJ_analysis}
Events with two \eLJs are selected by the logical or of single and di-electron triggers.
As mentioned in Section~\ref{sec:muLJ_analysis}, the electrons used for the \eLJ reconstruction are required to be within a \delr$=0.2$ cone around the trigger seed. Additional requirements on their \pt, identification and isolation are also applied, with tighter requirements than those applied online. In events selected by di-electron trigger, the two online electrons are required to match the two different \eLJs.

This analysis channel is optimised for the scenarios where the mass of the dark photon is smaller than the mass of a muon pair. To improve the sensitivity in this mass range, all the \eLJ candidates are required to be formed by two electrons with merged EM clusters.
In addition, the \LJs are required to be back-to-back, with a minimum separation in the azimuthal angle of at least 2.5 radians. Events are retained only if the mass imbalance of the two \eLJs is below 0.8.

The main background is composed by the random overlap of prompt electrons, produced in $Z$ boson decays or $t\bar{t}$ events, with additional ID tracks.
In order to reduce the contribution of \ttbar processes, selected events are rejected if at least one hadronic jet with $\pt>40\,\gev$ is present. Moreover, to remove events where one $Z$ boson decays into two electrons, the invariant mass of the system of the two \eLJs ($m(\eLJ,\eLJ)$) must be outside of the interval between $80$ and $100\,\gev$.
The SR is finally defined by requiring \ptimb to be smaller than 0.8 for the leading \eLJ, while \rphi has to be smaller than 0.96 for the farthest \eLJ.
The distributions of these two variables are shown in Figure~\ref{fig:eLJ_abcdvars}.

\begin{figure}[tb!]
\centering
\subfloat[]{\includegraphics[width=0.45\textwidth]{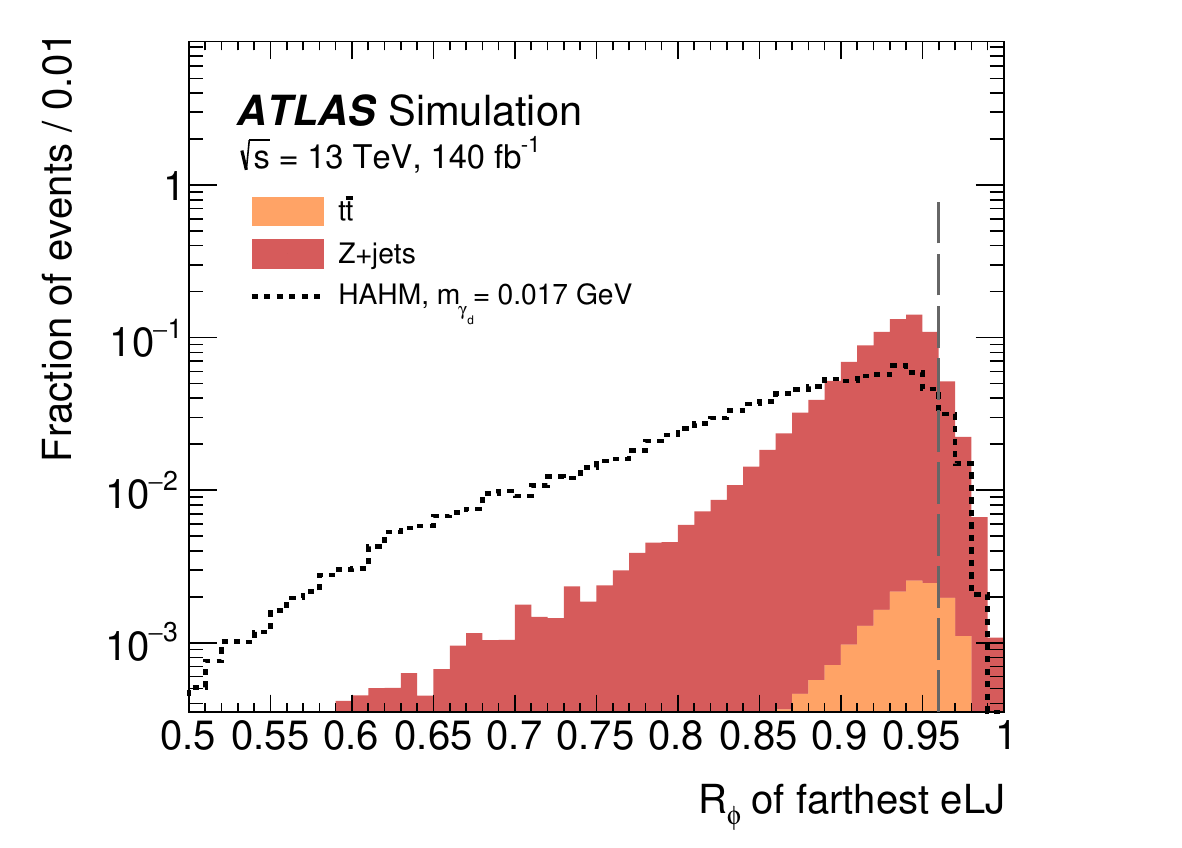}}
\subfloat[]{\includegraphics[width=0.45\textwidth]{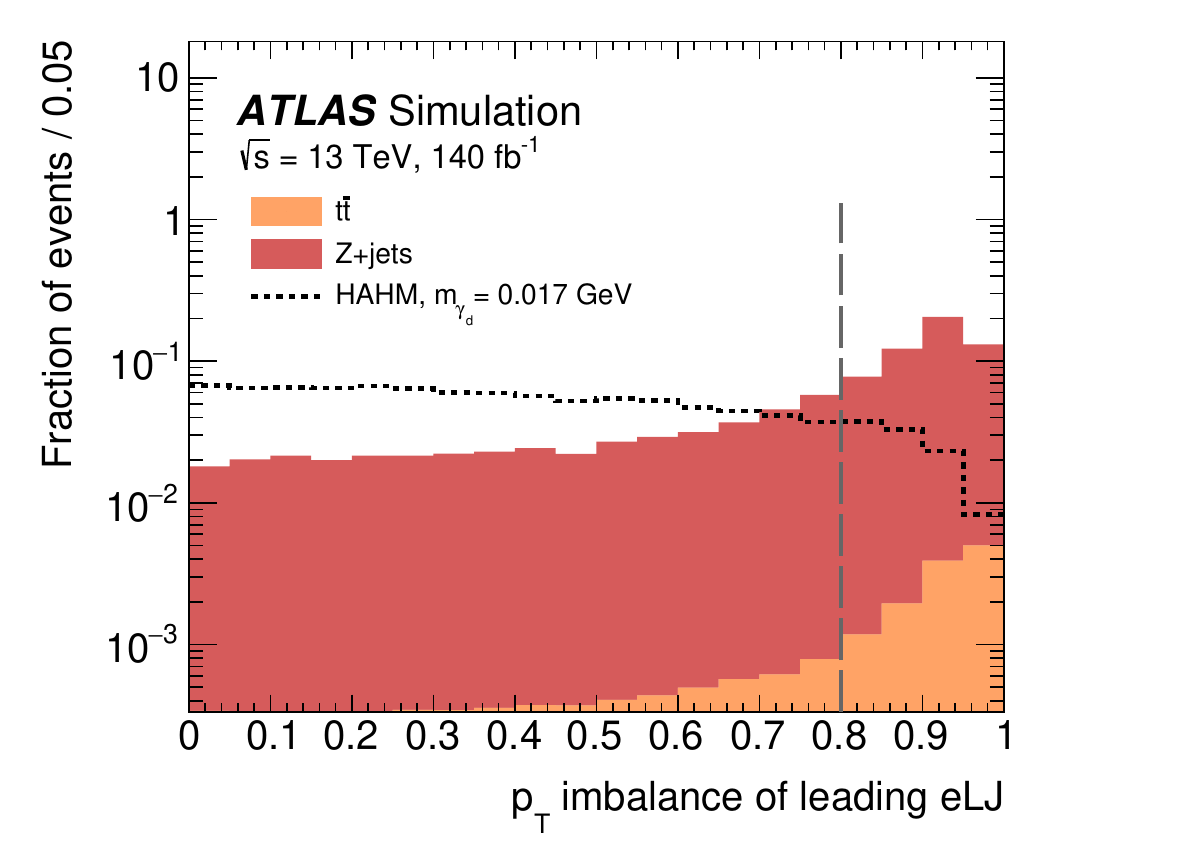}}
\caption{Distribution of the (a) \eLJ \rphi and (b) \eLJ \pt imbalance, obtained for signal events generated with the HAHM model with a \darky mass set to $17\,\mev$ and for simulated background events. Events are selected from the electron channel, with only the trigger requirement applied. Signal and background distributions are normalised to unit area. The dashed grey line corresponds to the selection applied in the definition of the SR.}
\label{fig:eLJ_abcdvars}
\end{figure}

After these selections, the residual background is estimated directly from data using the so-called \emph{ABCD} method.
The overlapping electron showers in the calorimeter for the low-mass range considered in this region prevent a well-defined \eLJ mass reconstruction and the use of the bump-hunting method, as in the muon channel.
In the plane defined by the \ptimb and \rphi variables, the four regions A, B, C and D, are identified:
region A corresponds to the SR, while regions B, C, D correspond to three CRs defined by reversing the selection on \rphi, \ptimb, and both the variables, respectively.
The requirements on the \ptimb and \rphi adopted in the definition of the SR are determined as the one that maximise the signal-to-background sensitivity, while minimising the signal contamination in regions B, C and D.
These variables are considered uncorrelated when \ptimb is considered for the leading eLJ and \rphi for the sub-leading eLJ, with \rphi being a calorimeter-based variable and \ptimb being track-based.

The number of background events in the SR ($N_A$) can be estimated by $N_A=(N_B\times N_C)/N_D$, where $N_i$ with $i={B,C,D}$ is the number of observed data events in the three CRs. This relation is extended using a likelihood-based method analogous to the one used in Ref.~\cite{EXOT-2022-15}, which fits the number of background and signal events simultaneously, taking into account any potential signal contamination in the CRs.

This method for the background estimate is validated in regions where the signal contribution is expected to be small. A first set of Validation Regions (VRs) is defined by combining CRs C and D, and CRs B and D, while another region ($\text{VR}_Z$) is defined by reversing the selection on the invariant mass of the two \eLJs in the event, hence targeting pairs of \LJs originating in $Z\rightarrow e^+e^-$ events.
The Pearson linear correlation coefficient between the two variables defining the ABCD plane in all control and validation regions was observed to be below 2\%.
The signal leakage in these regions was found to be less than 10\% of the total signal in the ABCD plane for all signal scenarios considered in the analysis.

The definitions of the SR A, CRs B, C and D, as well as the $\text{VR}_Z$ are summarised in Table~\ref{tab:electronSR}.
\begin{table}[!tb]
\caption{\label{tab:electronSR} Definition of the analysis regions of the electron channel.}
\def\arraystretch{1.2}
\begin{center}
\begin{tabular}{l r r r r r }
\toprule
Requirement / Region               & SR        & CR B & CR C & CR D & $\text{VR}_Z$  \\
\midrule
Applied to both leading and farthest \eLJ & & & & & \\
\midrule
Number of EM clusters in \eLJ     & \multicolumn{5}{c}{$1$} \\
\eLJ mass imbalance           & \multicolumn{5}{c}{$<0.8$} \\
\midrule
Selection on event-level variables & & & & & \\
\midrule
$\Delta\phi(e\text{LJ},e\text{LJ})$             & \multicolumn{5}{c}{$>2.5$} \\
Number of jets ($\pt > 40\,\gev$)       & \multicolumn{5}{c}{0} \\
$m(e\text{LJ},e\text{LJ}) \notin[80,100]\,\gev$ & yes    & yes    & yes    & yes    & veto   \\
Leading \eLJ \ptimb                 & $< 0.8$  & $<0.8$ & $> 0.8$ & $> 0.8$  & --  \\
Farthest \eLJ \rphi                 & $< 0.96$  & $>0.96$ & $< 0.96$ & $> 0.96$  & -- \\
\bottomrule
\end{tabular}
\end{center}
\end{table}

For each VR, alternative A, B, C, D regions are defined by dividing the plane in increasing steps of \ptimb and \rphi, in order to assert the reliability of the method in each of these regions.
In all the tested regions, the observed number of events is found to be in agreement with the expected value within one standard deviation.

Figure~\ref{fig:eLJ_abcd} shows the distribution of events in the ABCD planes, for observed data and events simulated with the HAHM model assuming a mass of the \darky of $17\,\mev$ and a decay branching ratio of the Higgs boson to dark sector particles of $0.5\%$. The number of observed data events in the \eLJ ABCD plane is reported in Table~\ref{tab:eLJdata}, where the yields in SR A are extracted by a background-only fit performed on data in all ABCD regions.
\begin{figure}[tb!]
\centering
\subfloat[]{\includegraphics[width=0.45\textwidth]{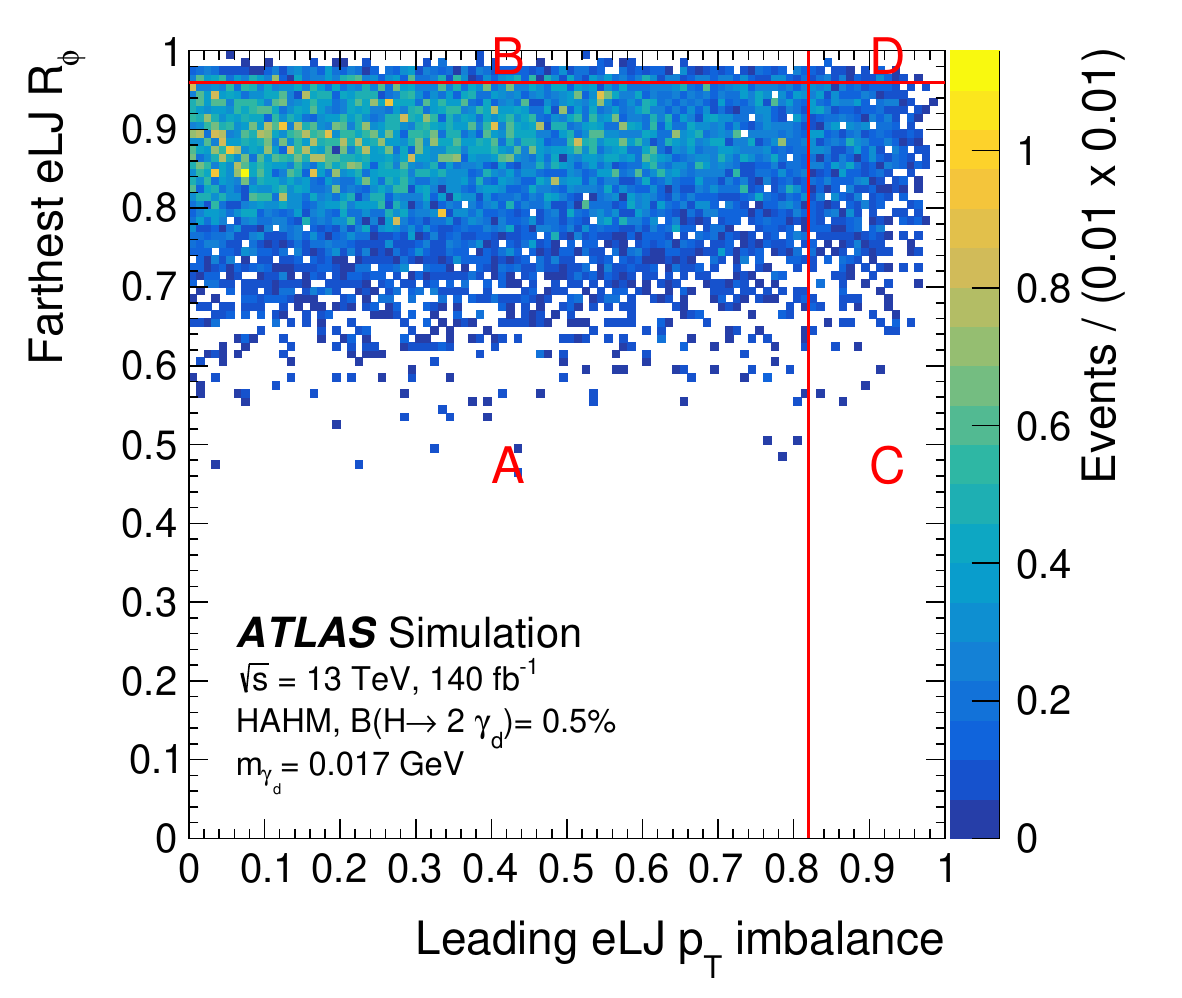}}
\subfloat[]{\includegraphics[width=0.45\textwidth]{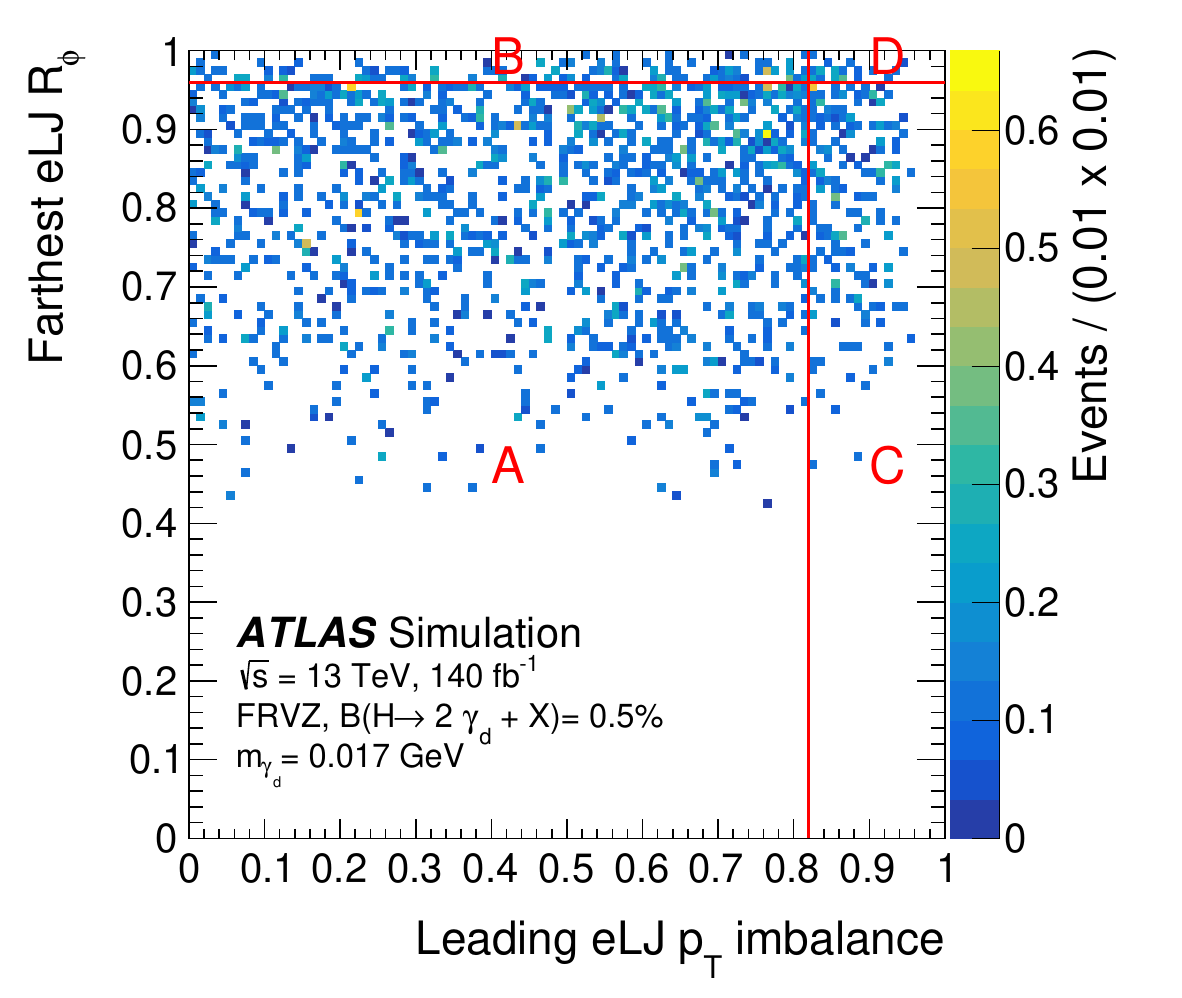}}\\
\subfloat[]{\includegraphics[width=0.45\textwidth]{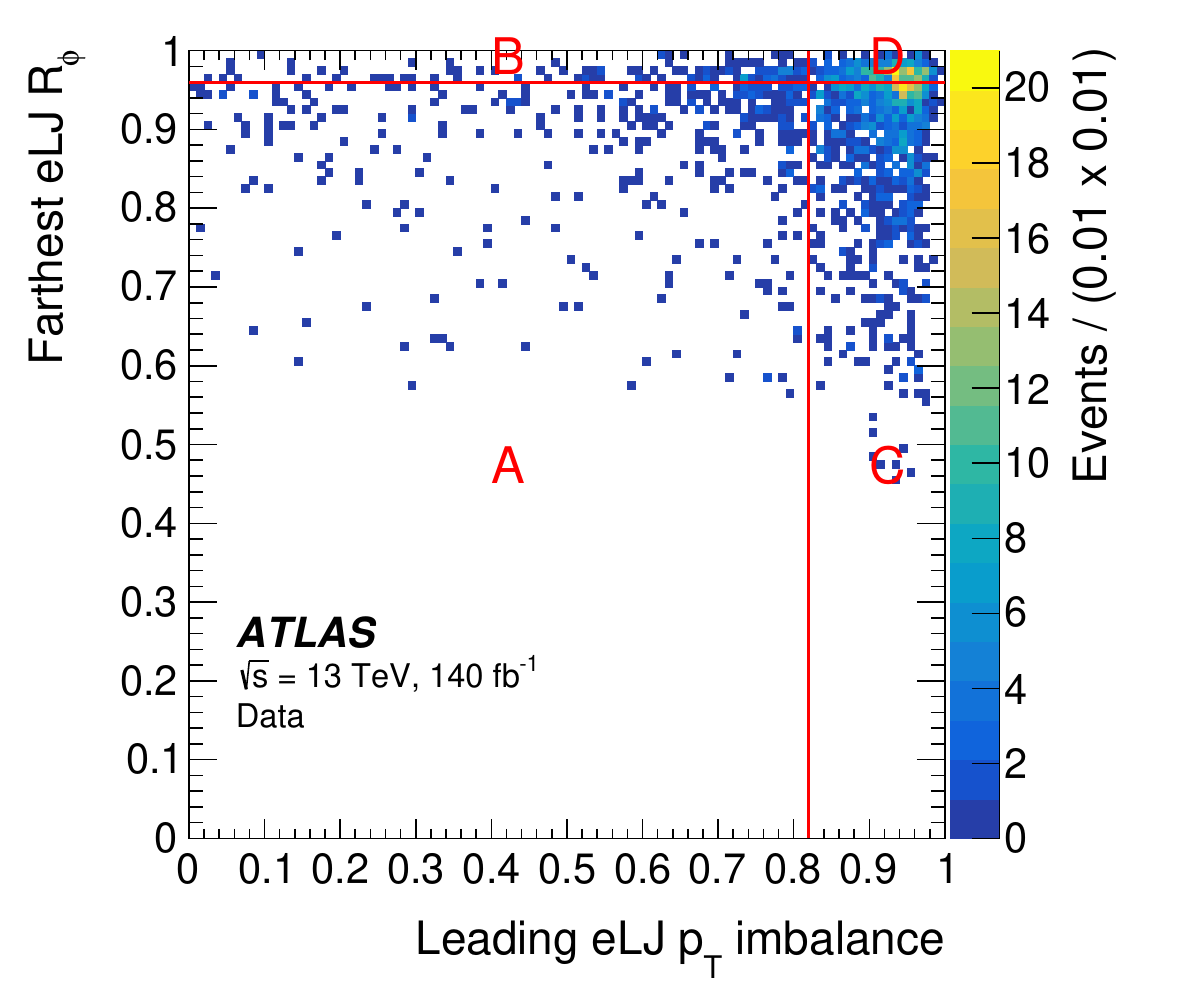}}
\caption{Distribution of (a, b) expected signal and (c) observed data events in the ABCD regions defined by the leading \eLJ \ptimb and farthest \eLJ \rphi variables in the electron channel. Signal events are simulated according to the (a) HAHM and (b) FRVZ models, for a \darky mass of $17\,\mev$ and assuming a branching ratio of the Higgs boson decay to dark photons of 0.5\%.}
\label{fig:eLJ_abcd}
\end{figure}

\begin{table}[tb!]
\caption{Observed data events in the four ABCD regions and the expected yields in the SR A were extracted with a background-only fit of the electron channel, assuming no signal.
The \textit{a priori} scenario does not consider data yields in SR A, whereas the \textit{a posteriori} scenario includes data in all ABCD regions in the fit.
The expected number of events is reported together with its statistical uncertainty.}
\label{tab:eLJdata}
\def\arraystretch{1.2}
\centering
\begin{tabular}{ccccccc}
\hline
Region    & CR B    & CR C    & CR D      & SR expected \textit{a priori}  & SR expected \textit{a posteriori}   & SR observed \\
\hline
\eLJ--\eLJ   & 125    & 862    & 356      & $303 \pm 33$   & $334 \pm 17$   & 351         \\
\hline
\end{tabular}
\end{table}


\FloatBarrier
%

%

\section{Systematic uncertainties}
\label{sec:uncertainties}
The overall uncertainty in the SRs yields is dominated by the background statistical uncertainty.
For the electron channel, the uncertainty in the observed yields in the ABCD control regions are propagated to the SR expectation obtained from the ABCD method, amounting to 5\%.
Other potential sources of experimental uncertainties are considered for the background estimates and the simulated signal yields.

The following experimental uncertainties are taken into account in the signal distributions for the \mLJ, \eLJ, and \mixLJ channels, as well as in the background estimations. All systematic uncertainties are incorporated into the fits using nuisance parameters, which are constrained by Gaussian terms in the likelihood function.

The background parameterisation in the muon channels is data-driven and extracted from the fit in the control region, thus only systematic uncertainties related to the background modelling and the fit are considered. The bias from the choice of the background model is evaluated with the spurious signal method, as discussed in Section~\ref{sec:muLJ_analysis}. At most three spurious signal events are extracted across the full spectrum and this number is taken as an uncertainty on the signal yields independent of the assumed dark photon mass.

To account for potential biases in the signal model description, an injection test is performed to determine if the correct number of events can be extracted using a signal-plus-background fit. Pseudo-data generated from the background model with a known number of injected signal events is used and a fit is performed. The obtained value for signal yields is then compared with the injected value. The largest difference per mass point between the injected and fitted signal for each of the configurations is assigned as a systematic uncertainty, which ranges from 1\% and up to 5\% increasing with the mass of the dark photon.

The uncertainty in the integrated luminosity of the combined data samples from 2015 to 2018 is 0.83\%, calculated using the methodology described in Ref.~\cite{DAPR-2021-01}. The uncertainty is determined using the LUCID-2 detector~\cite{LUCID2} for primary luminosity measurements, supplemented by measurements using the inner detector and calorimeters.

The pile-up modelling uncertainty that accounts for the difference between the simulated and measured inelastic $pp$ cross-section~\cite{STDM-2015-05} is evaluated with a data-to-MC re-weighting method of the distribution of the average number of interactions per bunch crossing. This uncertainty is propagated through the event selection, and results in a less than 4\% effect on the event yield of all the signal samples.

The experimental uncertainties are related to the lepton reconstruction, identification and isolation, and are evaluated using $Z\rightarrow \ell\ell$, $J/\Psi\rightarrow \ell\ell$ events in data and MC~\cite{EGAM-2018-01,MUON-2018-03}. The dominant uncertainties arise from muon isolation and electron identification found to be less than 1\% and 2\% for \mLJ and \eLJ channels respectively. The impact of electron isolation, lepton momentum resolution and energy scale, and muon identification are found to be sub-dominant. No additional systematic uncertainty is considered due to the additional correction added on top of the standard isolation variables.
The experimental uncertainties related to the jet energy resolution and jet energy scale, evaluated from the standard calibration scheme~\cite{PERF-2016-04}, amount to up to 3\% of the expected signal yields. This uncertainty is relevant only for the \twoeLJ channel where jets are used in the selection.

Uncertainty in the efficiency of trigger influence the expected yields and are estimated to be less than 2.2\% across all signal samples. The estimates are based on studies of the detector performance using Run 2 data and changes to the MC simulation~\cite{TRIG-2018-01,TRIG-2018-05}.

A summary of the experimental systematic uncertainties considered for the signal samples in this search is presented in Figure~\ref{fig:sysPlot}. The average uncertainties represent the bulk of the signal samples, with minor variations, typically of a few percent, observed as a function of the \darky mass.

\begin{figure}[tb!]
\centering
\includegraphics[width=.95\textwidth]{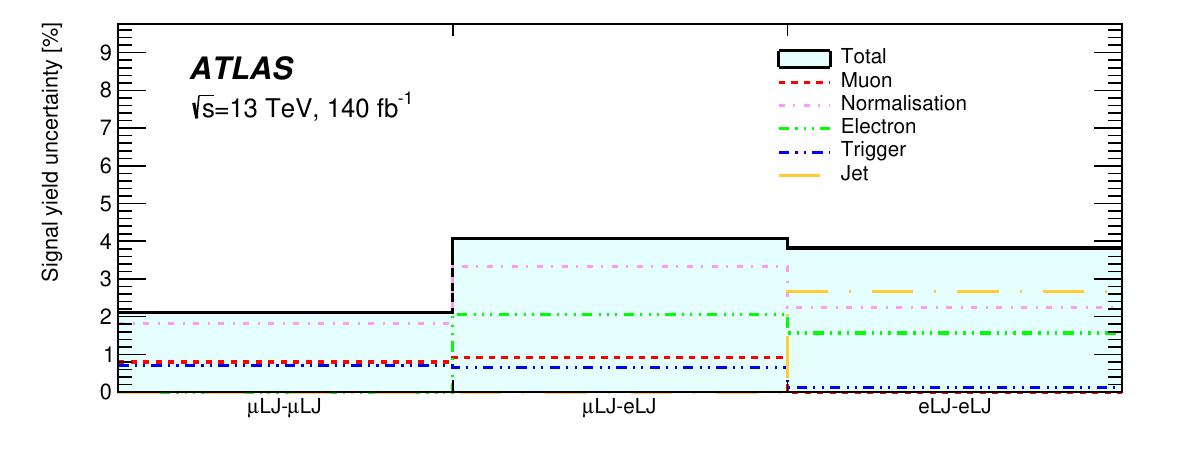}
\caption{
Contributions from the different sources of uncertainty to the signal yields in the search channels over all simulated signal samples. The reported uncertainties are averaged over all signal MC. The `Muon' and `Electron' sources contain all lepton-related systematic uncertainties and are dominated by the uncertainty in the muon isolation and electron identification. The `Triggers' source contains all trigger systematic uncertainties. The `Jet' source, relevant only for the electron channel, contains the jet energy scale and resolution uncertainties. The `Normalisation' source contains all systematic uncertainties affecting the overall normalisation of the yields.
}
\label{fig:sysPlot}
\end{figure}

\FloatBarrier


\FloatBarrier
%

%

%
\section{Results and interpretations}
\label{sec:result}
The statistical analysis of the muon channel uses an unbinned maximum-likelihood fit of the \mLJ mass distribution. The search is conducted in the \darky mass range between twice the muon mass and $20\,\gev$ for the HAHM model, while this range is limited up to $10\,\gev$ for the FRVZ model. The fit is performed with a step size of 10~\mev, smaller than the invariant mass resolution of the reconstructed \LJ.
Instead, the electron channel uses a binned maximum-likelihood fit to the signal region (A) and the three control regions (B, C and D), describing the ABCD constraint and taking into account any possible signal contamination in the control regions. For the electron channel, limits are computed for the \darky masses available from signal simulated samples (ranging from $0.017\,\gev$ to $0.4\,\gev$), as no parameterisation of the signal model is performed.

No significant excesses are observed, thus expected and observed 95\% confidence level (CL) exclusion limits on the cross-section times branching ratio of the process \Htwogd are computed as a function of the dark photon mass using the CL$_s$~\cite{Read:2002hq} modified frequentist approach. The exclusion limits are computed using the asymptotic approximation~\cite{Cowan:2010js}.
The validity of the asymptotic approximation is evaluated by comparing it with a full calculation using pseudo-experiments. The CL$_\text{s}$ values from both methods agree within 2\%, with the largest discrepancy being 10\% in the high mass region.
The 95\% CL upper limits for the HAHM and FRVZ models are depicted in Figure~\ref{fig:allLimits} for the muon and electron search channels.
The muon channel sets limit for \darky masses larger than twice the muon mass, while the electron channel covers the complementary region, down to the  mass of the lightest signal sample available (0.017\,\gev).
The electron channel is not combined with the muon channel for \darky masses above twice the muon mass, given the dominant sensitivity of the muon channel in this range. The upper limit is not extracted in regions where background resonances are present.

\begin{figure}[tb!]
\centering
\subfloat[]{\includegraphics[width=.49\textwidth]{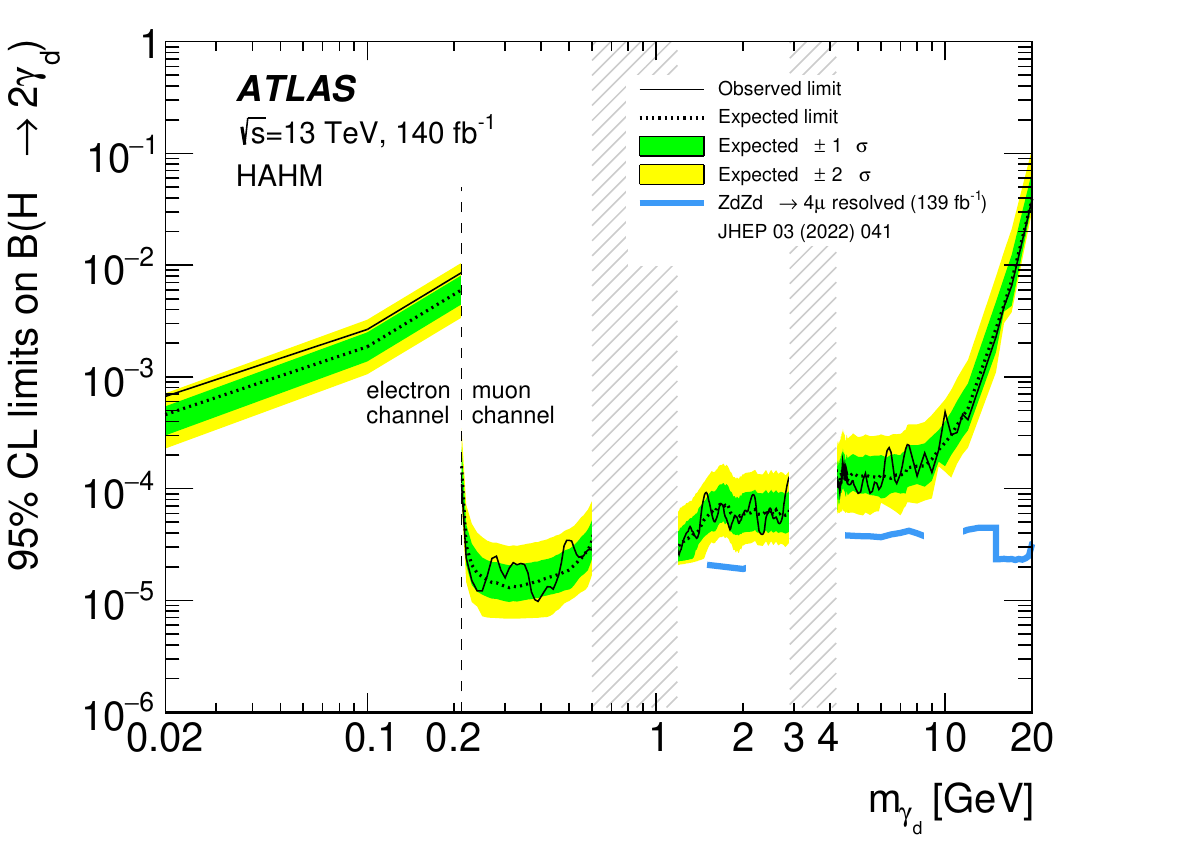}}
\subfloat[]{\includegraphics[width=.49\textwidth]{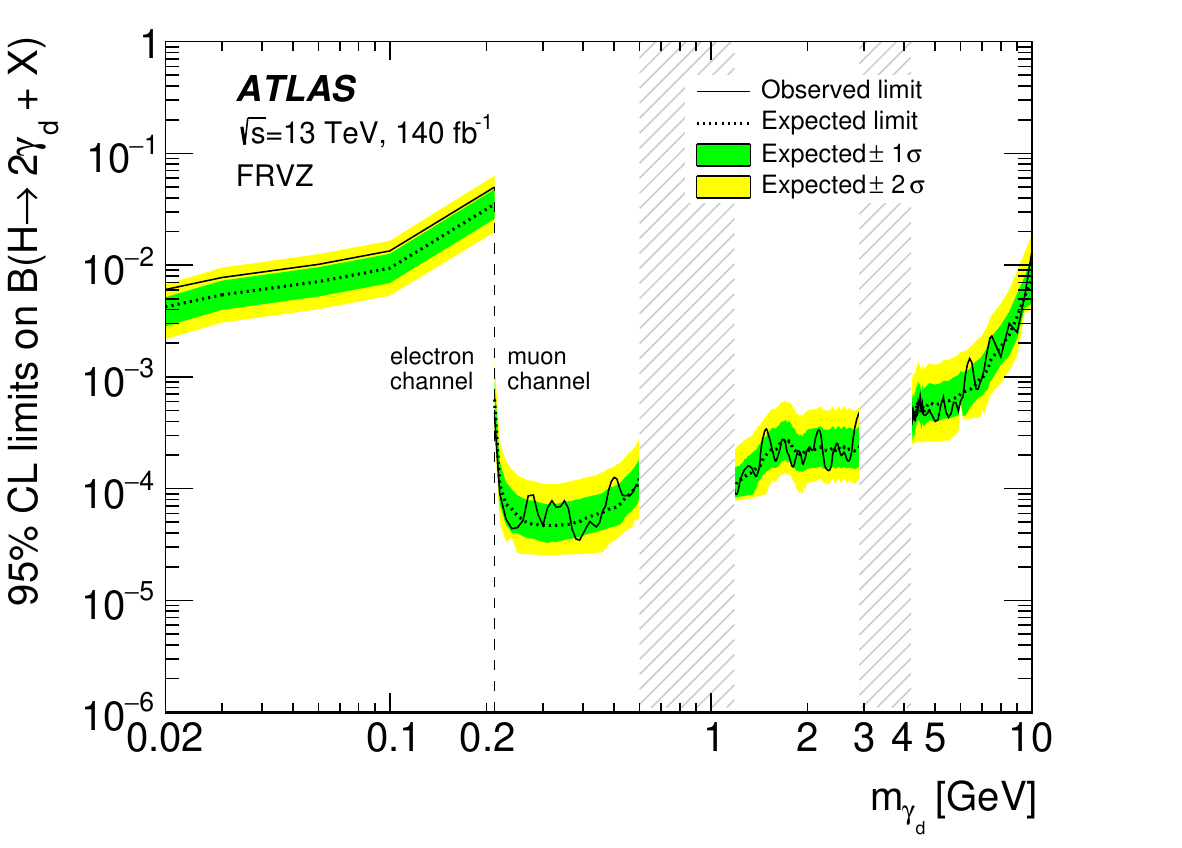}}
\caption{
95\% CL exclusion limits on the branching ratio of the Higgs boson to dark photons as a function of the \darky\ mass for the (a) HAHM and (b) FRVZ models.
The solid (dashed) black curve shows the observed (expected) exclusion limit and the green and yellow bands represent $\pm 1\sigma$ and $\pm 2\sigma$ uncertainty intervals around the expected limit.
The grey hatched regions indicate where the upper limit is not extracted.
The figure also shows the exclusion limit from the ATLAS search for resolved prompt \darky decays~\cite{HDBS-2018-55} (blue line).
}
\label{fig:allLimits}
\end{figure}

The search excludes a range of \Htwogd branching ratios from 0.001\% to 5\%, depending on the assumed dark photon mass, improving the previous search by ATLAS during Run~1~\cite{EXOT-2014-09} by a factor of 50 when considering a dark photon with a mass of 400~\mev. This improvement is mainly due to the renewed background estimate based on the invariant mass shape fit.
The muon channels are valid only for dark photon masses greater than twice the muon mass and complement the electron channel, which is covered only by ATLAS.
In the FRVZ scenarios, limits are derived down to $17\,\mev$ for the first time, significantly extending previous CMS and ATLAS searches for prompt \LJs.
In the HAHM scenarios, where the Higgs boson decays directly into a pair of dark photons, the sensitivity improves compared to the FRVZ model due to the harder dark photon energy spectrum.
In the high-mass range, when the muons from the dark-photon decay have large angular separation, making lepton-jet reconstruction inefficient, the analysis steeply loses sensitivity.
The result is complementary to the ATLAS search for four resolved prompt muons~\cite{HDBS-2018-55}, which sets stronger limits at larger masses and remains competitive around $\mdarky=2\,\gev$.
These are the first results for this model in ATLAS with the prompt \LJ signature.

Figure~\ref{fig:2Dlimitplot} presents the 90\% CL upper limits as a function of the kinetic mixing parameter $\epsilon$, which determines the lifetime of the dark photon, and \darky mass in both the HAHM vector-portal and FRVZ models, showing exclusion contours on the Higgs branching ratio to dark photons for branching fractions ranging from 0.01\% to 10\%.
The limits are interpolated between different masses by branching fraction variations~\cite{Batell:2009yf} as a function of the \darky mass, corrected by a linear interpolation of the signal efficiency between adjacent available MC signal samples.
Simple reweighting techniques~\cite{EXOT-2019-05} were applied to determine the analysis efficiency as a function of the lifetime for each mass point, using additional generated samples with varied lifetimes.
Exclusion regions from the complementary ATLAS search for displaced dark-photon~\cite{EXOT-2019-05} are also shown (orange line).
\begin{figure}[tb!]
\centering
\subfloat[]{\includegraphics[width=.5\textwidth]{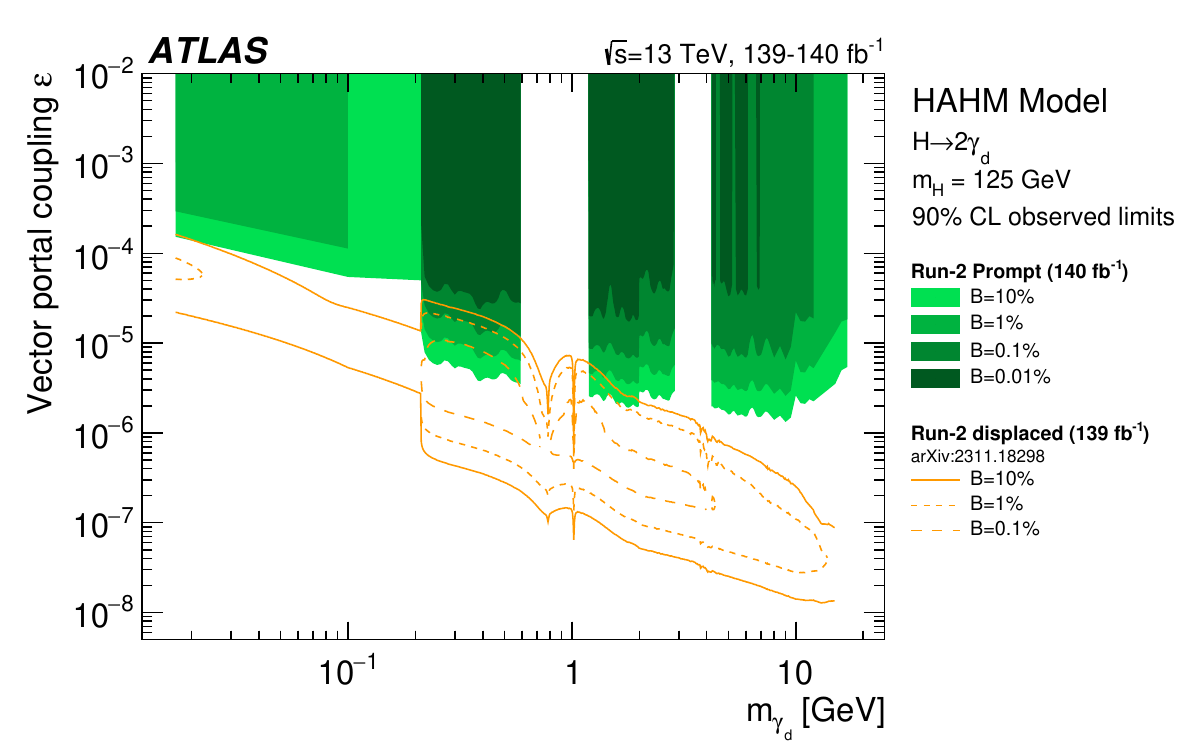}}
\subfloat[]{\includegraphics[width=.5\textwidth]{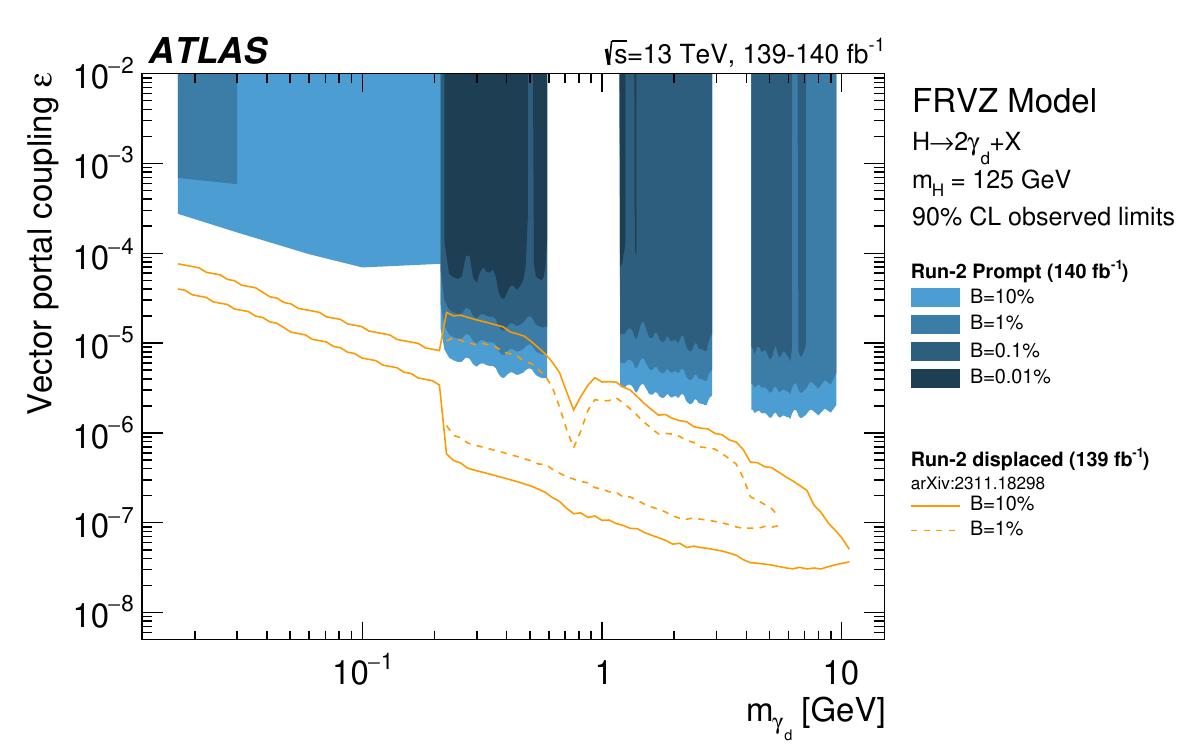}}
\caption{
90\% CL exclusion limit contours on the branching ratio of the Higgs boson to dark photons as functions of the \darky mass and kinetic mixing parameter $\epsilon$ for the (a) HAHM model and (b) FRVZ model.
These exclusions assume Higgs boson branching fractions to dark photons within a 0.01\% and 10\% range.
Excluded regions from the complementary ATLAS displaced dark-photon search~\cite{EXOT-2019-05} are displayed (orange lines) for comparison.
}
\label{fig:2Dlimitplot}
\end{figure}


\FloatBarrier
%

%

%
\section{Conclusion}
\label{sec:conclusion}
This paper details the first Run~2 search for light neutral particles decaying into collimated pairs of muons or electrons with the ATLAS detector at the LHC. By analysing data corresponding to an integrated luminosity of 140~fb$^{-1}$ of $pp$ collisions at a centre-of-mass energy of 13~\tev, the study investigates dark photons from Higgs boson decays, with a mass between 17~\mev and 20~\gev, extending the range of the previous search.
The data is consistent with background predictions and 95\% confidence level upper limits are set on the branching ratio of the Higgs boson to dark photons, ranging from 0.004\% and 5\% for the FRVZ model and from 0.001\% to 1\% for the HAHM model, depending on the dark photon mass.
Compared with the previous search conducted by ATLAS during Run~1, for the FRVZ model with a dark photon with a mass of 0.4~\gev, the upper limit on the branching ratio of the Higgs boson to dark photons is improved by approximately a factor of 50. When accounting for the contributions from the increased integrated luminosity and cross-section at a higher centre-of-mass energy, this results in an improvement by a factor of about 13.
The sensitivity of the muon channel has improved due to resonance search approach adopted, compared to simple use of the total event counts in the Run~1 analysis.
This is the first search for prompt Lepton-Jets in the electron channel using Run~2 data of the LHC and sets the most stringent limits to date.
The limits are extended down to $17\,\mev$ in dark photon mass, significantly improving previous CMS and ATLAS results for prompt Lepton-Jets.


\FloatBarrier

\section*{Acknowledgements}

%
%

%
%

We thank CERN for the very successful operation of the LHC and its injectors, as well as the support staff at
CERN and at our institutions worldwide without whom ATLAS could not be operated efficiently.

The crucial computing support from all WLCG partners is acknowledged gratefully, in particular from CERN, the ATLAS Tier-1 facilities at TRIUMF/SFU (Canada), NDGF (Denmark, Norway, Sweden), CC-IN2P3 (France), KIT/GridKA (Germany), INFN-CNAF (Italy), NL-T1 (Netherlands), PIC (Spain), RAL (UK) and BNL (USA), the Tier-2 facilities worldwide and large non-WLCG resource providers. Major contributors of computing resources are listed in Ref.~\cite{ATL-SOFT-PUB-2023-001}.

We gratefully acknowledge the support of ANPCyT, Argentina; YerPhI, Armenia; ARC, Australia; BMWFW and FWF, Austria; ANAS, Azerbaijan; CNPq and FAPESP, Brazil; NSERC, NRC and CFI, Canada; CERN; ANID, Chile; CAS, MOST and NSFC, China; Minciencias, Colombia; MEYS CR, Czech Republic; DNRF and DNSRC, Denmark; IN2P3-CNRS and CEA-DRF/IRFU, France; SRNSFG, Georgia; BMBF, HGF and MPG, Germany; GSRI, Greece; RGC and Hong Kong SAR, China; ISF and Benoziyo Center, Israel; INFN, Italy; MEXT and JSPS, Japan; CNRST, Morocco; NWO, Netherlands; RCN, Norway; MNiSW, Poland; FCT, Portugal; MNE/IFA, Romania; MSTDI, Serbia; MSSR, Slovakia; ARIS and MVZI, Slovenia; DSI/NRF, South Africa; MICIU/AEI, Spain; SRC and Wallenberg Foundation, Sweden; SERI, SNSF and Cantons of Bern and Geneva, Switzerland; NSTC, Taipei; TENMAK, T\"urkiye; STFC/UKRI, United Kingdom; DOE and NSF, United States of America.

Individual groups and members have received support from BCKDF, CANARIE, CRC and DRAC, Canada; CERN-CZ, FORTE and PRIMUS, Czech Republic; COST, ERC, ERDF, Horizon 2020, ICSC-NextGenerationEU and Marie Sk{\l}odowska-Curie Actions, European Union; Investissements d'Avenir Labex, Investissements d'Avenir Idex and ANR, France; DFG and AvH Foundation, Germany; Herakleitos, Thales and Aristeia programmes co-financed by EU-ESF and the Greek NSRF, Greece; BSF-NSF and MINERVA, Israel; NCN and NAWA, Poland; La Caixa Banking Foundation, CERCA Programme Generalitat de Catalunya and PROMETEO and GenT Programmes Generalitat Valenciana, Spain; G\"{o}ran Gustafssons Stiftelse, Sweden; The Royal Society and Leverhulme Trust, United Kingdom.

In addition, individual members wish to acknowledge support from Armenia: Yerevan Physics Institute (FAPERJ); CERN: European Organization for Nuclear Research (CERN PJAS); Chile: Agencia Nacional de Investigaci\'on y Desarrollo (FONDECYT 1230812, FONDECYT 1230987, FONDECYT 1240864); China: Chinese Ministry of Science and Technology (MOST-2023YFA1605700, MOST-2023YFA1609300), National Natural Science Foundation of China (NSFC - 12175119, NSFC 12275265, NSFC-12075060); Czech Republic: Czech Science Foundation (GACR - 24-11373S), Ministry of Education Youth and Sports (FORTE CZ.02.01.01/00/22\_008/0004632), PRIMUS Research Programme (PRIMUS/21/SCI/017); EU: H2020 European Research Council (ERC - 101002463); European Union: European Research Council (ERC - 948254, ERC 101089007), Horizon 2020 Framework Programme (MUCCA - CHIST-ERA-19-XAI-00), European Union, Future Artificial Intelligence Research (FAIR-NextGenerationEU PE00000013), Italian Center for High Performance Computing, Big Data and Quantum Computing (ICSC, NextGenerationEU); France: Agence Nationale de la Recherche (ANR-20-CE31-0013, ANR-21-CE31-0013, ANR-21-CE31-0022, ANR-22-EDIR-0002), Investissements d'Avenir Labex (ANR-11-LABX-0012); Germany: Baden-Württemberg Stiftung (BW Stiftung-Postdoc Eliteprogramme), Deutsche Forschungsgemeinschaft (DFG - 469666862, DFG - CR 312/5-2); Italy: Istituto Nazionale di Fisica Nucleare (ICSC, NextGenerationEU), Ministero dell'Università e della Ricerca (PRIN - 20223N7F8K - PNRR M4.C2.1.1); Japan: Japan Society for the Promotion of Science (JSPS KAKENHI JP22H01227, JSPS KAKENHI JP22H04944, JSPS KAKENHI JP22KK0227, JSPS KAKENHI JP23KK0245); Netherlands: Netherlands Organisation for Scientific Research (NWO Veni 2020 - VI.Veni.202.179); Norway: Research Council of Norway (RCN-314472); Poland: Ministry of Science and Higher Education (IDUB AGH, POB8, D4 no 9722), Polish National Agency for Academic Exchange (PPN/PPO/2020/1/00002/U/00001), Polish National Science Centre (NCN 2021/42/E/ST2/00350, NCN OPUS nr 2022/47/B/ST2/03059, NCN UMO-2019/34/E/ST2/00393, NCN \& H2020 MSCA 945339, UMO-2020/37/B/ST2/01043, UMO-2021/40/C/ST2/00187, UMO-2022/47/O/ST2/00148, UMO-2023/49/B/ST2/04085, UMO-2023/51/B/ST2/00920); Slovenia: Slovenian Research Agency (ARIS grant J1-3010); Spain: Generalitat Valenciana (Artemisa, FEDER, IDIFEDER/2018/048), Ministry of Science and Innovation (MCIN \& NextGenEU PCI2022-135018-2, MICIN \& FEDER PID2021-125273NB, RYC2019-028510-I, RYC2020-030254-I, RYC2021-031273-I, RYC2022-038164-I), PROMETEO and GenT Programmes Generalitat Valenciana (CIDEGENT/2019/027); Sweden: Carl Trygger Foundation (Carl Trygger Foundation CTS 22:2312), Swedish Research Council (Swedish Research Council 2023-04654, VR 2018-00482, VR 2022-03845, VR 2022-04683, VR 2023-03403, VR grant 2021-03651), Knut and Alice Wallenberg Foundation (KAW 2018.0157, KAW 2018.0458, KAW 2019.0447, KAW 2022.0358); Switzerland: Swiss National Science Foundation (SNSF - PCEFP2\_194658); United Kingdom: Leverhulme Trust (Leverhulme Trust RPG-2020-004), Royal Society (NIF-R1-231091); United States of America: U.S. Department of Energy (ECA DE-AC02-76SF00515), Neubauer Family Foundation.

%
%


%

%
\clearpage
\printbibliography
\clearpage
\input{atlas_authlist}
%

\end{document}

%% file: atlas_authlist.tex
 
\begin{flushleft}
\hypersetup{urlcolor=black}
{\Large The ATLAS Collaboration}

\bigskip

\AtlasOrcid[0000-0002-6665-4934]{G.~Aad}$^\textrm{\scriptsize 104}$,
\AtlasOrcid[0000-0001-7616-1554]{E.~Aakvaag}$^\textrm{\scriptsize 17}$,
\AtlasOrcid[0000-0002-5888-2734]{B.~Abbott}$^\textrm{\scriptsize 123}$,
\AtlasOrcid[0000-0002-0287-5869]{S.~Abdelhameed}$^\textrm{\scriptsize 119a}$,
\AtlasOrcid[0000-0002-1002-1652]{K.~Abeling}$^\textrm{\scriptsize 56}$,
\AtlasOrcid[0000-0001-5763-2760]{N.J.~Abicht}$^\textrm{\scriptsize 50}$,
\AtlasOrcid[0000-0002-8496-9294]{S.H.~Abidi}$^\textrm{\scriptsize 30}$,
\AtlasOrcid[0009-0003-6578-220X]{M.~Aboelela}$^\textrm{\scriptsize 45}$,
\AtlasOrcid[0000-0002-9987-2292]{A.~Aboulhorma}$^\textrm{\scriptsize 36e}$,
\AtlasOrcid[0000-0001-5329-6640]{H.~Abramowicz}$^\textrm{\scriptsize 155}$
% ...
\bigskip
\\

% ...
$^{17}$Department for Physics and Technology, University of Bergen, Bergen; Norway.\\
% ...
$^{30}$Physics Department, Brookhaven National Laboratory, Upton NY; United States of America.\\
% ...
$^{36}$$^{(a)}$Facult\'e des Sciences Ain Chock, Universit\'e Hassan II de Casablanca;$^{(b)}$Facult\'{e} des Sciences, Universit\'{e} Ibn-Tofail, K\'{e}nitra;$^{(c)}$Facult\'e des Sciences Semlalia, Universit\'e Cadi Ayyad, LPHEA-Marrakech;$^{(d)}$LPMR, Facult\'e des Sciences, Universit\'e Mohamed Premier, Oujda;$^{(e)}$Facult\'e des sciences, Universit\'e Mohammed V, Rabat;$^{(f)}$Institute of Applied Physics, Mohammed VI Polytechnic University, Ben Guerir; Morocco.\\
% ...
$^{45}$Physics Department, Southern Methodist University, Dallas TX; United States of America.\\
% ...
$^{50}$Fakult\"{a}t Physik , Technische Universit{\"a}t Dortmund, Dortmund; Germany.\\
% ...
$^{56}$II. Physikalisches Institut, Georg-August-Universit\"{a}t G\"ottingen, G\"ottingen; Germany.\\
% ...
$^{104}$CPPM, Aix-Marseille Universit\'e, CNRS/IN2P3, Marseille; France.\\
% ...
$^{119}$$^{(a)}$New York University Abu Dhabi, Abu Dhabi;$^{(b)}$United Arab Emirates University, Al Ain; United Arab Emirates.\\
% ...
$^{123}$Homer L. Dodge Department of Physics and Astronomy, University of Oklahoma, Norman OK; United States of America.\\
% ...
$^{155}$Raymond and Beverly Sackler School of Physics and Astronomy, Tel Aviv University, Tel Aviv; Israel.\\
% ...
$^{a}$ Also Affiliated with an institute covered by a cooperation agreement with CERN.\\
$^{b}$ Also at An-Najah National University, Nablus; Palestine.\\
$^{c}$ Also at Borough of Manhattan Community College, City University of New York, New York NY; United States of America.\\
$^{d}$ Also at Center for High Energy Physics, Peking University; China.\\
$^{e}$ Also at Center for Interdisciplinary Research and Innovation (CIRI-AUTH), Thessaloniki; Greece.\\
$^{f}$ Also at CERN, Geneva; Switzerland.\\
% ...

\end{flushleft}


%% file: ANA-EXOT-2018-55-PAPER.bib
@article{Dokshitzer:1997in,
    author = "Dokshitzer, Yuri L. and Leder, G. D. and Moretti, S. and Webber, B. R.",
    title = "{Better jet clustering algorithms}",
    eprint = "hep-ph/9707323",
    archivePrefix = "arXiv",
    reportNumber = "CAVENDISH-HEP-97-06",
    doi = "10.1088/1126-6708/1997/08/001",
    journal = "JHEP",
    volume = "08",
    pages = "001",
    year = "1997"
}

@article{ArkaniHamed:2008qp,
  author        = {Arkani-Hamed, Nima and Weiner, Neal},
  title         = {{LHC signals for a SuperUnified theory of Dark Matter}},
  journal       = {JHEP},
  volume        = {12},
  year          = {2008},
  pages         = {104},
  doi           = {10.1088/1126-6708/2008/12/104},
  eprint        = {0810.0714},
  archiveprefix = {arXiv},
  primaryclass  = {hep-ph},
  slaccitation  = {%%CITATION = ARXIV:0810.0714;%%}
}

@article{Baumgart:2009tn,
  author        = {Baumgart, Matthew and Cheung, Clifford and Ruderman,
                   Joshua T. and Wang, Lian-Tao and Yavin, Itay},
  title         = {{Non-abelian dark sectors and their collider signatures}},
  journal       = {JHEP},
  volume        = {04},
  year          = {2009},
  pages         = {014},
  doi           = {10.1088/1126-6708/2009/04/014},
  eprint        = {0901.0283},
  archiveprefix = {arXiv},
  primaryclass  = {hep-ph},
  slaccitation  = {%%CITATION = ARXIV:0901.0283;%%}
}

@article{Falkowski:2014ffa,
  author        = {Falkowski, Adam and Vega-Morales, Roberto},
  title         = {{Exotic Higgs decays in the golden channel}},
  journal       = {JHEP},
  volume        = {12},
  year          = {2014},
  pages         = {037},
  doi           = {10.1007/JHEP12(2014)037},
  eprint        = {1405.1095},
  archiveprefix = {arXiv},
  primaryclass  = {hep-ph},
  reportnumber  = {LPT-ORSAY-14-23},
  slaccitation  = {%%CITATION = ARXIV:1405.1095;%%}
}

@article{Curtin:2014cca,
  author        = {Curtin, David and Essig, Rouven and Gori, Stefania and
                   Shelton, Jessie},
  title         = {{Illuminating dark photons with high-energy colliders}},
  journal       = {JHEP},
  volume        = {02},
  year          = {2015},
  pages         = {157},
  doi           = {10.1007/JHEP02(2015)157},
  eprint        = {1412.0018},
  archiveprefix = {arXiv},
  primaryclass  = {hep-ph},
  reportnumber  = {YITP-SB-14-49},
  slaccitation  = {%%CITATION = ARXIV:1412.0018;%%}
}

@article{Falkowski:2010cm,
  author        = {Falkowski, Adam and Ruderman, Joshua T. and Volansky,
                   Tomer and Zupan, Jure},
  title         = {{Hidden Higgs decaying to lepton jets}},
  journal       = {JHEP},
  volume        = {05},
  year          = {2010},
  pages         = {077},
  doi           = {10.1007/JHEP05(2010)077},
  eprint        = {1002.2952},
  archiveprefix = {arXiv},
  primaryclass  = {hep-ph},
  slaccitation  = {%%CITATION = ARXIV:1002.2952;%%}
}

@article{Falkowski:2010gv,
  author        = {Falkowski, Adam and Ruderman, Joshua T. and Volansky,
                   Tomer and Zupan, Jure},
  title         = {{Discovering Higgs Boson Decays to Lepton Jets at Hadron
                   Colliders}},
  journal       = {Phys. Rev. Lett.},
  volume        = {105},
  year          = {2010},
  pages         = {241801},
  doi           = {10.1103/PhysRevLett.105.241801},
  eprint        = {1007.3496},
  archiveprefix = {arXiv},
  primaryclass  = {hep-ph},
  slaccitation  = {%%CITATION = ARXIV:1007.3496;%%}
}

@article{Batell:2009yf,
  author        = {Batell, Brian and Pospelov, Maxim and Ritz, Adam},
  title         = {{Probing a secluded U(1) at B factories}},
  journal       = {Phys. Rev. D},
  volume        = {79},
  year          = {2009},
  pages         = {115008},
  doi           = {10.1103/PhysRevD.79.115008},
  eprint        = {0903.0363},
  archiveprefix = {arXiv},
  primaryclass  = {hep-ph},
  slaccitation  = {%%CITATION = ARXIV:0903.0363;%%}
}

@article{Aaij:2015tna,
  author        = {{LHCb Collaboration}},
  title         = {{Search for Hidden-Sector Bosons in $B^0 \!\to K^{*0}\mu^+\mu^-$ Decays}},
  collaboration = {LHCb},
  journal       = {Phys. Rev. Lett.},
  volume        = {115},
  year          = {2015},
  number        = {16},
  pages         = {161802},
  doi           = {10.1103/PhysRevLett.115.161802},
  eprint        = {1508.04094},
  archiveprefix = {arXiv},
  primaryclass  = {hep-ex},
  reportnumber  = {CERN-PH-EP-2015-202, LHCB-PAPER-2015-036},
  slaccitation  = {%%CITATION = ARXIV:1508.04094;%%}
}

@article{Blumlein:2011mv,
  author        = {Bl{\"u}mlein, Johannes and Brunner, Jurgen},
  title         = {{New exclusion limits for dark gauge forces from beam-dump data}},
  journal       = {Phys. Lett. B},
  volume        = {701},
  year          = {2011},
  pages         = {155-159},
  doi           = {10.1016/j.physletb.2011.05.046},
  eprint        = {1104.2747},
  archiveprefix = {arXiv},
  primaryclass  = {hep-ex},
  reportnumber  = {DESY-11-062, DO-TH-11-11, SFB-CPP-11-18, LPN-11-17,
                   DESY-11--062, DO--TH-11-11, SFB-CPP--11--18, LPN-11--17},
  slaccitation  = {%%CITATION = ARXIV:1104.2747;%%}
}

@article{Bjorken:2009mm,
  author        = {Bjorken, James D. and Essig, Rouven and Schuster, Philip
                   and Toro, Natalia},
  title         = {{New fixed-target experiments to search for dark gauge forces}},
  journal       = {Phys. Rev. D},
  volume        = {80},
  year          = {2009},
  pages         = {075018},
  doi           = {10.1103/PhysRevD.80.075018},
  eprint        = {0906.0580},
  archiveprefix = {arXiv},
  primaryclass  = {hep-ph},
  reportnumber  = {SLAC-PUB-13650, SU-ITP-09-22},
  slaccitation  = {%%CITATION = ARXIV:0906.0580;%%}
}

@article{Bross:1989mp,
  author       = {Bross, A. and Crisler, M. and Pordes, Stephen H. and
                  Volk, J. and Errede, S. and Wrbanek, J.},
  title        = {{Search for short-lived particles produced in an electron beam dump}},
  journal      = {Phys. Rev. Lett.},
  volume       = {67},
  year         = {1991},
  pages        = {2942-2945},
  doi          = {10.1103/PhysRevLett.67.2942},
  reportnumber = {FERMILAB-PUB-89-138-E},
  slaccitation = {%%CITATION = PRLTA,67,2942;%%}
}

@article{Merkel:2011ze,
  author        = {{A1 Collaboration}},
  title         = {{Search for Light Gauge Bosons of the Dark Sector at the
                   Mainz Microtron}},
  collaboration = {A1},
  journal       = {Phys. Rev. Lett.},
  volume        = {106},
  year          = {2011},
  pages         = {251802},
  doi           = {10.1103/PhysRevLett.106.251802},
  eprint        = {1101.4091},
  archiveprefix = {arXiv},
  primaryclass  = {nucl-ex},
  slaccitation  = {%%CITATION = ARXIV:1101.4091;%%}
}

@article{Adlarson:2013eza,
  author        = {{WASA-at-COSY Collaboration}},
  title         = {{Search for a dark photon in the $\pi^0 \to e^+e^-\gamma$
                   decay}},
  collaboration = {WASA-at-COSY},
  journal       = {Phys. Lett. B},
  volume        = {726},
  year          = {2013},
  pages         = {187-193},
  doi           = {10.1016/j.physletb.2013.08.055},
  eprint        = {1304.0671},
  archiveprefix = {arXiv},
  primaryclass  = {hep-ex},
  slaccitation  = {%%CITATION = ARXIV:1304.0671;%%}
}

@article{Abrahamyan:2011gv,
    author = "Abrahamyan, S. and others",
    collaboration = "APEX",
    title = "{Search for a New Gauge Boson in Electron-Nucleus Fixed-Target Scattering by the APEX Experiment}",
    eprint = "1108.2750",
    archivePrefix = "arXiv",
    primaryClass = "hep-ex",
    reportNumber = "JLAB-PHY-11-1406, SLAC-PUB-14491",
    doi = "10.1103/PhysRevLett.107.191804",
    journal = "Phys. Rev. Lett.",
    volume = "107",
    pages = "191804",
    year = "2011"
}

@article{Reece:2009un,
  author        = {Reece, Matthew and Wang, Lian-Tao},
  title         = {{Searching for the light dark gauge boson in GeV-scale
                   experiments}},
  journal       = {JHEP},
  volume        = {07},
  year          = {2009},
  pages         = {051},
  doi           = {10.1088/1126-6708/2009/07/051},
  eprint        = {0904.1743},
  archiveprefix = {arXiv},
  primaryclass  = {hep-ph},
  slaccitation  = {%%CITATION = ARXIV:0904.1743;%%}
}

@article{Blumlein:2013cua,
  author        = {Bl{\"u}mlein, Johannes and Brunner, Jorgen},
  title         = {{New exclusion limits on dark gauge forces from proton Bremsstrahlung in beam-dump data}},
  journal       = {Phys. Lett. B},
  volume        = {731},
  year          = {2014},
  pages         = {320-326},
  doi           = {10.1016/j.physletb.2014.02.029},
  eprint        = {1311.3870},
  archiveprefix = {arXiv},
  primaryclass  = {hep-ph},
  reportnumber  = {DESY-13-202, DO-TH-13-29, SFB-CPP-13-87, LPN-13-087},
  slaccitation  = {%%CITATION = ARXIV:1311.3870;%%}
}

@article{Gninenko:2012eq,
  author        = {Gninenko, S. N.},
  title         = {{Constraints on sub-GeV hidden sector gauge bosons from a
                   search for heavy neutrino decays}},
  journal       = {Phys. Lett. B},
  volume        = {713},
  year          = {2012},
  pages         = {244-248},
  doi           = {10.1016/j.physletb.2012.06.002},
  eprint        = {1204.3583},
  archiveprefix = {arXiv},
  primaryclass  = {hep-ph},
  slaccitation  = {%%CITATION = ARXIV:1204.3583;%%}
}

@article{Essig:2010gu,
  author        = {Essig, Rouven and Harnik, Roni and Kaplan, Jared and
                   Toro, Natalia},
  title         = {{Discovering new light states at neutrino experiments}},
  journal       = {Phys. Rev. D},
  volume        = {82},
  year          = {2010},
  pages         = {113008},
  doi           = {10.1103/PhysRevD.82.113008},
  eprint        = {1008.0636},
  archiveprefix = {arXiv},
  primaryclass  = {hep-ph},
  reportnumber  = {SLAC-PUB-14197, FERMILAB-PUB-10-274-T},
  slaccitation  = {%%CITATION = ARXIV:1008.0636;%%}
}

@article{Agakishiev:2013fwl,
  author        = {{HADES Collaboration}},
  title         = {{Searching a dark photon with HADES}},
  collaboration = {HADES},
  journal       = {Phys. Lett. B},
  volume        = {731},
  year          = {2014},
  pages         = {265-271},
  doi           = {10.1016/j.physletb.2014.02.035},
  eprint        = {1311.0216},
  archiveprefix = {arXiv},
  primaryclass  = {hep-ex},
  slaccitation  = {%%CITATION = ARXIV:1311.0216;%%}
}

@article{Archilli:2011zc,
  author        = {{KLOE-2 Collaboration}},
  title         = {{Search for a vector gauge boson in $\phi$ meson decays
                   with the KLOE detector}},
  collaboration = {KLOE-2},
  journal       = {Phys. Lett. B},
  volume        = {706},
  year          = {2012},
  pages         = {251-255},
  doi           = {10.1016/j.physletb.2011.11.033},
  eprint        = {1110.0411},
  archiveprefix = {arXiv},
  primaryclass  = {hep-ex},
  slaccitation  = {%%CITATION = ARXIV:1110.0411;%%}
}

@article{Babusci:2012cr,
  author        = {{KLOE-2 Collaboration}},
  title         = {{Limit on the production of a light vector gauge boson in
                   $\phi$ meson decays with the KLOE detector}},
  collaboration = {KLOE-2},
  journal       = {Phys. Lett. B},
  volume        = {720},
  year          = {2013},
  pages         = {111-115},
  doi           = {10.1016/j.physletb.2013.01.067},
  eprint        = {1210.3927},
  archiveprefix = {arXiv},
  primaryclass  = {hep-ex},
  slaccitation  = {%%CITATION = ARXIV:1210.3927;%%}
}

@article{Aubert:2009cp,
  author        = {{BABAR Collaboration}},
  title         = {{Search for Dimuon Decays of a Light Scalar Boson in
                   Radiative Transitions $\Upsilon \to \gamma A^0$}},
  collaboration = {BaBar},
  journal       = {Phys. Rev. Lett.},
  volume        = {103},
  year          = {2009},
  pages         = {081803},
  doi           = {10.1103/PhysRevLett.103.081803},
  eprint        = {0905.4539},
  archiveprefix = {arXiv},
  primaryclass  = {hep-ex},
  reportnumber  = {BABAR-PUB-09-015, SLAC-PUB-13653},
  slaccitation  = {%%CITATION = ARXIV:0905.4539;%%}
}

@article{Lees:2014xha,
  author        = {{BABAR Collaboration}},
  title         = {{Search for a Dark Photon in $e^+e^-$ Collisions at
                   BaBar}},
  collaboration = {BaBar},
  journal       = {Phys. Rev. Lett.},
  volume        = {113},
  year          = {2014},
  number        = {20},
  pages         = {201801},
  doi           = {10.1103/PhysRevLett.113.201801},
  eprint        = {1406.2980},
  archiveprefix = {arXiv},
  primaryclass  = {hep-ex},
  reportnumber  = {BABAR-PUB-14-002, SLAC-PUB-15979},
  slaccitation  = {%%CITATION = ARXIV:1406.2980;%%}
}

@article{Lees:2015rxq,
  author        = {{BABAR Collaboration}},
  title         = {{Search for Long-Lived Particles in $e^+e^-$ Collisions}},
  collaboration = {BaBar},
  journal       = {Phys. Rev. Lett.},
  volume        = {114},
  year          = {2015},
  number        = {17},
  pages         = {171801},
  doi           = {10.1103/PhysRevLett.114.171801},
  eprint        = {1502.02580},
  archiveprefix = {arXiv},
  primaryclass  = {hep-ex},
  reportnumber  = {BABAR-PUB-14-010, SLAC-PUB-16226},
  slaccitation  = {%%CITATION = ARXIV:1502.02580;%%}
}

@article{TheBelle:2015mwa,
  author        = {{Belle Collaboration}},
  title         = {{Search for the Dark Photon and the Dark Higgs Boson at Belle}},
  collaboration = {Belle},
  journal       = {Phys. Rev. Lett.},
  volume        = {114},
  year          = {2015},
  number        = {21},
  pages         = {211801},
  doi           = {10.1103/PhysRevLett.114.211801},
  eprint        = {1502.00084},
  archiveprefix = {arXiv},
  primaryclass  = {hep-ex},
  reportnumber  = {Belle-Preprint-2014-20, KEK-Preprint-2014-36},
  slaccitation  = {%%CITATION = ARXIV:1502.00084;%%}
}

@article{Won:2016pjz,
  author        = {{Belle Collaboration}},
  title         = {{Search for a dark vector gauge boson decaying to $\pi^+
                   \pi^-$ using $\eta \rightarrow \pi^+\pi^- \gamma$ decays}},
  collaboration = {Belle},
  journal       = {Phys. Rev. D},
  volume        = {94},
  year          = {2016},
  number        = {9},
  pages         = {092006},
  doi           = {10.1103/PhysRevD.94.092006},
  eprint        = {1609.05599},
  archiveprefix = {arXiv},
  primaryclass  = {hep-ex},
  slaccitation  = {%%CITATION = ARXIV:1609.05599;%%}
}

@article{Ablikim:2018bhf,
  author        = {{BESIII Collaboration}},
  title         = {{Measurement of $\mathcal{B}(J/\psi \to \eta' e^+ e^-)$
                   and search for a dark photon}},
  collaboration = {BESIII},
  journal       = {Phys. Rev. D},
  volume        = {99},
  year          = {2019},
  number        = {1},
  pages         = {012013},
  doi           = {10.1103/PhysRevD.99.012013},
  eprint        = {1809.00635},
  archiveprefix = {arXiv},
  primaryclass  = {hep-ex},
  slaccitation  = {%%CITATION = ARXIV:1809.00635;%%}
}

@article{Ablikim:2017aab,
  author        = {{BESIII Collaboration}},
  title         = {{Dark photon search in the mass range between 1.5 and 3.4 GeV/$c^2$}},
  collaboration = {BESIII},
  journal       = {Phys. Lett. B},
  volume        = {774},
  year          = {2017},
  pages         = {252-257},
  doi           = {10.1016/j.physletb.2017.09.067},
  eprint        = {1705.04265},
  archiveprefix = {arXiv},
  primaryclass  = {hep-ex},
  slaccitation  = {%%CITATION = ARXIV:1705.04265;%%}
}

@article{Pospelov:2008zw,
  author        = {Pospelov, Maxim},
  title         = {{Secluded U(1) below the weak scale}},
  journal       = {Phys. Rev. D},
  volume        = {80},
  year          = {2009},
  pages         = {095002},
  doi           = {10.1103/PhysRevD.80.095002},
  eprint        = {0811.1030},
  archiveprefix = {arXiv},
  primaryclass  = {hep-ph},
  slaccitation  = {%%CITATION = ARXIV:0811.1030;%%}
}

@article{Davoudiasl:2012ig,
  author        = {Davoudiasl, Hooman and Lee, Hye-Sung and Marciano,
                   William J.},
  title         = {{Dark side of Higgs diphoton decays and muon $g - 2$}},
  journal       = {Phys. Rev. D},
  volume        = {86},
  year          = {2012},
  pages         = {095009},
  doi           = {10.1103/PhysRevD.86.095009},
  eprint        = {1208.2973},
  archiveprefix = {arXiv},
  primaryclass  = {hep-ph},
  slaccitation  = {%%CITATION = ARXIV:1208.2973;%%}
}

@article{Endo:2012hp,
  author        = {Endo, Motoi and Hamaguchi, Koichi and Mishima, Go},
  title         = {{Constraints on hidden photon models from electron $g - 2$
                   and hydrogen spectroscopy}},
  journal       = {Phys. Rev. D},
  volume        = {86},
  year          = {2012},
  pages         = {095029},
  doi           = {10.1103/PhysRevD.86.095029},
  eprint        = {1209.2558},
  archiveprefix = {arXiv},
  primaryclass  = {hep-ph},
  reportnumber  = {UT-12-31, IPMU12-172},
  slaccitation  = {%%CITATION = ARXIV:1209.2558;%%}
}

@article{Chang:2018rso,
  author        = {Chang, Jae Hyeok and Essig, Rouven and McDermott, Samuel D.},
  title         = {{Supernova 1987A constraints on sub-GeV dark sectors, millicharged particles, the QCD axion, and an axion-like particle}},
  eprint        = {1803.00993},
  archiveprefix = {arXiv},
  primaryclass  = {hep-ph},
  reportnumber  = {YITP-SB-18-01, FERMILAB-PUB-17-432-T},
  doi           = {10.1007/JHEP09(2018)051},
  journal       = {JHEP},
  volume        = {09},
  pages         = {051},
  year          = {2018}
}

@article{Dreiner:2013mua,
  author        = {Dreiner, Herbert K. and Fortin, Jean-Francois and
                   Hanhart, Christoph and Ubaldi, Lorenzo},
  title         = {{Supernova constraints on MeV dark sectors from $e^+e^-$
                   annihilations}},
  journal       = {Phys. Rev. D},
  volume        = {89},
  year          = {2014},
  number        = {10},
  pages         = {105015},
  doi           = {10.1103/PhysRevD.89.105015},
  eprint        = {1310.3826},
  archiveprefix = {arXiv},
  primaryclass  = {hep-ph},
  reportnumber  = {BONN-TH-2013-18, CERN-PH-TH-2013-242, SU-ITP-13-19,
                   CETUP2013-017},
  slaccitation  = {%%CITATION = ARXIV:1310.3826;%%}
}

@article{Cepeda:2019klc,
  author        = {Cepeda, M. and others},
  editor        = {Dainese, Andrea and Mangano, Michelangelo and Meyer, Andreas B. and Nisati, Aleandro and Salam, Gavin and Vesterinen, Mika Anton},
  title         = {{Higgs Physics at the HL-LHC and HE-LHC}},
  eprint        = {1902.00134},
  archiveprefix = {arXiv},
  primaryclass  = {hep-ph},
  reportnumber  = {CERN-LPCC-2018-04},
  doi           = {10.23731/CYRM-2019-007.221},
  journal       = {CERN Yellow Rep. Monogr.},
  volume        = {7},
  pages         = {221--584},
  year          = {2019}
}

@article{LHCb:2019vmc,
  author        = {{LHCb Collaboration}},
  collaboration = {LHCb},
  title         = {{Search for $A'\to\mu^+\mu^-$ Decays}},
  eprint        = {1910.06926},
  archiveprefix = {arXiv},
  primaryclass  = {hep-ex},
  reportnumber  = {LHCb-PAPER-2019-031, CERN-EP-2019-212},
  doi           = {10.1103/PhysRevLett.124.041801},
  journal       = {Phys. Rev. Lett.},
  volume        = {124},
  number        = {4},
  pages         = {041801},
  year          = {2020}
}

@mastersthesis{Oreglia:1980cs,
    author = "Oreglia, M.",
    title = "{A Study of the Reactions $\psi^\prime \to \gamma \gamma \psi$}",
    reportNumber = "SLAC-0236, SLAC-236, UMI-81-08973, SLAC-R-0236, SLAC-R-236",
    type = "PhD Thesis",
    month = "12",
    year = "1980",
    url = "http://www-public.slac.stanford.edu/sciDoc/docMeta.aspx?slacPubNumber=slac-r-236.html",
}

@article{Abel:2008ai,
    author = "Abel, S. A. and Goodsell, M. D. and Jaeckel, J. and Khoze, V. V. and Ringwald, A.",
    title = "{Kinetic Mixing of the Photon with Hidden U(1)s in String Phenomenology}",
    eprint = "0803.1449",
    archivePrefix = "arXiv",
    primaryClass = "hep-ph",
    reportNumber = "IPPP-08-14, DESY-08-026",
    doi = "10.1088/1126-6708/2008/07/124",
    journal = "JHEP",
    volume = "07",
    pages = "124",
    year = "2008"
}

@article{CMS:2024qxz, 
    author = "{CMS Collaboration}", 
    title = "{Search for long-lived particles decaying to final states with a pair of muons in proton-proton collisions at $ \sqrt{s} $ = 13.6 TeV}", 
    eprint = "2402.14491", 
    archivePrefix = "arXiv", 
    primaryClass = "hep-ex", 
    reportNumber = "CMS-EXO-23-014, CERN-EP-2024-025", 
    doi = "10.1007/JHEP05(2024)047", 
    journal = "JHEP", 
    volume = "05", 
    pages = "047", 
    year = "2024" 

}

@article{FASER:2024-plb,
title = {Search for dark photons with the FASER detector at the LHC},
journal = {Physics Letters B},
volume = {848},
pages = {138378},
year = {2024},
doi = {https://doi.org/10.1016/j.physletb.2023.138378},
author = "{FASER Collaboration}"
}

@article{KLOE2:2015-plb,
title = {Search for dark Higgsstrahlung in e$^+$e$^\rightarrow\mu^+\mu^-$ and missing energy events with the KLOE experiment},
journal = {Physics Letters B},
volume = {747},
pages = {365-372},
year = {2015},
issn = {0370-2693},
doi = {https://doi.org/10.1016/j.physletb.2015.06.015},
author = "{KLOE-2 Collaboration}"
}

@article{BABAR:PRL2012,
  title = {Search for Low-Mass Dark-Sector Higgs Bosons},
  author        = {{BABAR Collaboration}},
  collaboration = {BaBar},
  journal = {Phys. Rev. Lett.},
  volume = {108},
  issue = {21},
  pages = {211801},
  numpages = {7},
  year = {2012},
  month = {05},
  publisher = {American Physical Society},
  doi = {10.1103/PhysRevLett.108.211801},
}

@article{Belle:2020,
    author = "{Belle Collaboration}",
    collaboration = "Belle",
    title = "{Search for the dark photon in $B^0 \to A^{\prime} A^{\prime}$, $A^{\prime} \to e^+ e^-$, $\mu^+ \mu^-$, and $\pi^+ \pi^-$ decays at Belle}",
    eprint = "2012.02538",
    archivePrefix = "arXiv",
    primaryClass = "hep-ex",
    reportNumber = "Belle Preprint 2020-19, KEK Preprint 2020-36",
    doi = "10.1007/JHEP04(2021)191",
    journal = "JHEP",
    volume = "04",
    pages = "191",
    year = "2021"
}

@article{Belle2:PRL2023,
  title = {Search for a Dark Photon and an Invisible Dark Higgs Boson in ${\ensuremath{\mu}}^{+}{\ensuremath{\mu}}^{\ensuremath{-}}$ and Missing Energy Final States with the Belle II Experiment},
  author = "{Belle II Collaboration}",
  collaboration = {Belle II Collaboration},
  journal = {Phys. Rev. Lett.},
  volume = {130},
  issue = {7},
  pages = {071804},
  numpages = {8},
  year = {2023},
  month = {Feb},
  publisher = {American Physical Society},
  doi = {10.1103/PhysRevLett.130.071804},
}


%% file: ATLAS-useful.bib
@Article{Cacciari:2008gp,
     author    = "Cacciari, Matteo and Salam, Gavin P. and Soyez, Gregory",
     title     = "{The anti-\(k_{t}\) jet clustering algorithm}",
     journal   = "JHEP",
     volume    = "04",
     year      = "2008",
     pages     = "063",
     eprint    = "0802.1189",
     archivePrefix = "arXiv",
     primaryClass  =  "hep-ph",
     doi       = "10.1088/1126-6708/2008/04/063",
     SLACcitation  = "%%CITATION = 0802.1189;%%"
}

@Article{Fastjet,
      author         = "Cacciari, Matteo and Salam, Gavin P. and Soyez, Gregory",
      title          = "{FastJet user manual}",
      journal        = "Eur. Phys. J. C",
      volume         = "72",
      year           = "2012",
      pages          = "1896",
      doi            = "10.1140/epjc/s10052-012-1896-2",
      eprint         = "1111.6097",
      archivePrefix  = "arXiv",
      primaryClass   = "hep-ph",
      reportNumber   = "CERN-PH-TH-2011-297",
      SLACcitation   = "%%CITATION = ARXIV:1111.6097;%%"
}

@article{Ball:2012cx,
      author         = "{NNPDF Collaboration} and Ball, Richard D. and others",
      title          = "{Parton distributions with LHC data}",
      journal        = "Nucl. Phys. B",
      volume         = "867",
      year           = "2013",
      pages          = "244",
      doi            = "10.1016/j.nuclphysb.2012.10.003",
      eprint         = "1207.1303",
      archivePrefix  = "arXiv",
      primaryClass   = "hep-ph",
      reportNumber   = "EDINBURGH-2012-08, IFUM-FT-997, FR-PHENO-2012-014,
                        RWTH-TTK-12-25, CERN-PH-TH-2012-037, SFB-CPP-12-47\,
                        --CERN-PH-TH-2012-037",
      SLACcitation   = "%%CITATION = ARXIV:1207.1303;%%"
}

@article{Ball:2014uwa,
      author         = "{NNPDF Collaboration} and Ball, Richard D. and others",
      title          = "{Parton distributions for the LHC run II}",
      journal        = "JHEP",
      volume         = "04",
      year           = "2015",
      pages          = "040",
      doi            = "10.1007/JHEP04(2015)040",
      eprint         = "1410.8849",
      archivePrefix  = "arXiv",
      primaryClass   = "hep-ph",
      reportNumber   = "EDINBURGH-2014-15, IFUM-1034-FT, CERN-PH-TH-2013-253,
                        OUTP-14-11P, CAVENDISH-HEP-14-11",
      SLACcitation   = "%%CITATION = ARXIV:1410.8849;%%"
}

@Article{Sjostrand:2007gs,
	author    = "Sj{\"o}strand, T. and Mrenna, S. and Skands, P.",
	title     = "{A brief introduction to PYTHIA 8.1}",
	journal   = "Comput. Phys. Commun.",
	volume    = "178",
	year      = "2008",
	pages     = "852-867",
	eprint    = "0710.3820",
	archivePrefix = "arXiv",
	primaryClass  =  "hep-ph",
	doi       = "10.1016/j.cpc.2008.01.036",
	SLACcitation  = "%%CITATION = 0710.3820;%%"
}

@article{Sjostrand:2014zea,
      author         = "Sj{\"o}strand, Torbj{\"o}rn and Ask, Stefan and Christiansen,
                        Jesper R. and Corke, Richard and Desai, Nishita and Ilten,
                        Philip and Mrenna, Stephen and Prestel, Stefan and
                        Rasmussen, Christine O. and Skands, Peter Z.",
      title          = "{An introduction to PYTHIA 8.2}",
      journal        = "Comput. Phys. Commun.",
      volume         = "191",
      year           = "2015",
      pages          = "159",
      doi            = "10.1016/j.cpc.2015.01.024",
      eprint         = "1410.3012",
      archivePrefix  = "arXiv",
      primaryClass   = "hep-ph",
      reportNumber   = "LU-TP-14-36, MCNET-14-22, CERN-PH-TH-2014-190,
                        FERMILAB-PUB-14-316-CD, DESY-14-178, SLAC-PUB-16122,
                        --FERMILAB-PUB-14-316-CD",
      SLACcitation   = "%%CITATION = ARXIV:1410.3012;%%"
}

@Article{Lange:2001uf,
      author         = "Lange, D. J.",
      title          = "{The EvtGen particle decay simulation package}",
      booktitle      = "{Proceedings, 7th International Conference on B physics
                        at hadron machines (BEAUTY 2000)}",
      journal        = "Nucl. Instrum. Meth. A",
      volume         = "462",
      year           = "2001",
      pages          = "152",
      doi            = "10.1016/S0168-9002(01)00089-4",
      SLACcitation   = "%%CITATION = NUIMA,A462,152;%%"
}

@Article{Alwall:2014hca,
      author         = "Alwall, J. and Frederix, R. and Frixione, S. and Hirschi,
                        V. and Maltoni, F. and Mattelaer, O. and Shao, H. -S. and
                        Stelzer, T. and Torrielli, P. and Zaro, M.",
      title          = "{The automated computation of tree-level and
                        next-to-leading order differential cross sections, and
                        their matching to parton shower simulations}",
      journal        = "JHEP",
      volume         = "07",
      year           = "2014",
      pages          = "079",
      doi            = "10.1007/JHEP07(2014)079",
      eprint         = "1405.0301",
      archivePrefix  = "arXiv",
      primaryClass   = "hep-ph",
      reportNumber   = "CERN-PH-TH-2014-064, CP3-14-18, LPN14-066, MCNET-14-09,
                        ZU-TH-14-14",
      SLACcitation   = "%%CITATION = ARXIV:1405.0301;%%"
}

@Article{Nason:2004rx,
      author         = "Nason, Paolo",
      title          = "{A new method for combining NLO QCD with shower Monte Carlo algorithms}",
      journal        = "JHEP",
      volume         = "11",
      pages          = "040",
      doi            = "10.1088/1126-6708/2004/11/040",
      year           = "2004",
      eprint         = "hep-ph/0409146",
      archivePrefix  = "arXiv",
}

@Article{Frixione:2007vw,
      author         = "Frixione, Stefano and Nason, Paolo and Oleari, Carlo",
      title          = "{Matching NLO QCD computations with parton shower
                        simulations: the POWHEG method}",
      journal        = "JHEP",
      volume         = "11",
      pages          = "070",
      doi            = "10.1088/1126-6708/2007/11/070",
      year           = "2007",
      eprint         = "0709.2092",
      archivePrefix  = "arXiv",
      primaryClass   = "hep-ph",
}

@Article{Alioli:2010xd,
      author         = "Alioli, Simone and Nason, Paolo and Oleari, Carlo and Re,
                        Emanuele",
      title          = "{A general framework for implementing NLO calculations in
                        shower Monte Carlo programs: the POWHEG BOX}",
      journal        = "JHEP",
      volume         = "06",
      pages          = "043",
      doi            = "10.1007/JHEP06(2010)043",
      year           = "2010",
      eprint         = "1002.2581",
      archivePrefix  = "arXiv",
      primaryClass   = "hep-ph",
}

@article{Anastasiou:2003ds,
      author         = "Anastasiou, Charalampos and Dixon, Lance and Melnikov,
                        Kirill and Petriello, Frank",
      title          = "{High-precision QCD at hadron colliders: Electroweak
                        gauge boson rapidity distributions at next-to-next-to leading order}",
      journal        = "Phys. Rev. D",
      volume         = "69",
      year           = "2004",
      pages          = "094008",
      doi            = "10.1103/PhysRevD.69.094008",
      eprint         = "hep-ph/0312266",
      archivePrefix  = "arXiv",
      reportNumber   = "SLAC-PUB-10288, UH-511-1042-03",
      SLACcitation   = "%%CITATION = HEP-PH/0312266;%%"
}

@article{Frixione:2007nw,
      author         = "Frixione, Stefano and Ridolfi, Giovanni and Nason, Paolo",
      title          = "{A positive-weight next-to-leading-order Monte Carlo for
                        heavy flavour hadroproduction}",
      journal        = "JHEP",
      volume         = "09",
      pages          = "126",
      doi            = "10.1088/1126-6708/2007/09/126",
      year           = "2007",
      eprint         = "0707.3088",
      archivePrefix  = "arXiv",
      primaryClass   = "hep-ph",
      SLACcitation   = "%%CITATION = ARXIV:0707.3088;%%",
}

@Article{Hoeche:2009rj,
      author         = "H{\"o}che, Stefan and Krauss, Frank and Schumann, Steffen and Siegert, Frank",
      title          = "{QCD matrix elements and truncated showers}",
      journal        = "JHEP",
      volume         = "05",
      pages          = "053",
      doi            = "10.1088/1126-6708/2009/05/053",
      year           = "2009",
      eprint         = "0903.1219",
      archivePrefix  = "arXiv",
      primaryClass   = "hep-ph",
}

@Article{Gleisberg:2008fv,
      author         = "Gleisberg, Tanju and H{\"o}che, Stefan",
      title          = "{Comix, a new matrix element generator}",
      journal        = "JHEP",
      volume         = "12",
      pages          = "039",
      doi            = "10.1088/1126-6708/2008/12/039",
      year           = "2008",
      eprint         = "0808.3674",
      archivePrefix  = "arXiv",
      primaryClass   = "hep-ph",
}

@Article{Schumann:2007mg,
      author         = "Schumann, Steffen and Krauss, Frank",
      title          = "{A parton shower algorithm based on Catani--Seymour dipole factorisation}",
      journal        = "JHEP",
      volume         = "03",
      pages          = "038",
      doi            = "10.1088/1126-6708/2008/03/038",
      year           = "2008",
      eprint         = "0709.1027",
      archivePrefix  = "arXiv",
      primaryClass   = "hep-ph",
}

@article{Hoeche:2012yf,
      author         = "H{\"o}che, Stefan and Krauss, Frank and Sch{\"o}nherr, Marek and
                        Siegert, Frank",
      title          = "{QCD matrix elements + parton showers. The NLO case}",
      journal        = "JHEP",
      volume         = "04",
      year           = "2013",
      pages          = "027",
      doi            = "10.1007/JHEP04(2013)027",
      eprint         = "1207.5030",
      archivePrefix  = "arXiv",
      primaryClass   = "hep-ph",
      reportNumber   = "SLAC-PUB-15191, IPPP-12-52, DCPT-12-104, LPN12-081,
                        FR-PHENO-2012-017, MCNET-12-09, --FR-PHENO-2012-017",
      SLACcitation   = "%%CITATION = ARXIV:1207.5030;%%"
}

@article{Bothmann:2019yzt,
      author = "Bothmann, Enrico and others",
      title = "{Event generation with Sherpa 2.2}",
      journal = "SciPost Phys.",
      volume = "7",
      year = "2019",
      number = "3",
      pages = "034",
      doi = "10.21468/SciPostPhys.7.3.034",
      reportNumber = "FERMILAB-PUB-19-218-T, SLAC-PUB-17433, IPPP/19/42, MCNET-19-11",
      eprint = "1905.09127",
      archivePrefix = "arXiv",
      primaryClass = "hep-ph",
}

@article{Hoeche:2011fd,
      author         = "H{\"o}che, Stefan and Krauss, Frank and Sch{\"o}nherr, Marek and
                        Siegert, Frank",
      title          = "{A critical appraisal of NLO+PS matching methods}",
      journal        = "JHEP",
      volume         = "09",
      year           = "2012",
      pages          = "049",
      doi            = "10.1007/JHEP09(2012)049",
      eprint         = "1111.1220",
      archivePrefix  = "arXiv",
      primaryClass   = "hep-ph",
      reportNumber   = "SLAC-PUB-14661, IPPP-11-67, DCPT-11-134, LPN11-58,
                        FR-PHENO-2011-019, MCNET-11-24",
      SLACcitation   = "%%CITATION = ARXIV:1111.1220;%%"
}

@article{Catani:2001cc,
      author         = "Catani, S. and Krauss, F. and Webber, B. R. and Kuhn, R.",
      title          = "{QCD Matrix Elements + Parton Showers}",
      journal        = "JHEP",
      volume         = "11",
      year           = "2001",
      pages          = "063",
      doi            = "10.1088/1126-6708/2001/11/063",
      eprint         = "hep-ph/0109231",
      archivePrefix  = "arXiv",
      reportNumber   = "CERN-TH-2000-367, CAVENDISH-HEP-00-03",
      SLACcitation   = "%%CITATION = HEP-PH/0109231;%%"
}

@article{Cascioli:2011va,
      author         = "Cascioli, Fabio and Maierh{\"o}fer, Philipp and Pozzorini, Stefano",
      title          = "{Scattering Amplitudes with Open Loops}",
      journal        = "Phys. Rev. Lett.",
      volume         = "108",
      year           = "2012",
      pages          = "111601",
      doi            = "10.1103/PhysRevLett.108.111601",
      eprint         = "1111.5206",
      archivePrefix  = "arXiv",
      primaryClass   = "hep-ph",
      reportNumber   = "ZU-TH-23-11, LPN11-66",
      SLACcitation   = "%%CITATION = ARXIV:1111.5206;%%"
}

@article{Buccioni:2019sur,
    author = "Buccioni, Federico and Lang, Jean-Nicolas and Lindert, Jonas M. and Maierh{\"o}fer, Philipp and Pozzorini, Stefano and Zhang, Hantian and Zoller, Max F.",
    title = "{OpenLoops 2}",
    eprint = "1907.13071",
    archivePrefix = "arXiv",
    primaryClass = "hep-ph",
    reportNumber = "IPPP/19/62, FR-PHENO-2019-12, PSI-PR-19-15, ZU-TH 37/19",
    doi = "10.1140/epjc/s10052-019-7306-2",
    journal = "Eur. Phys. J. C",
    volume = "79",
    number = "10",
    pages = "866",
    year = "2019"
}

@article{Denner:2016kdg,
      author         = "Denner, Ansgar and Dittmaier, Stefan and Hofer, Lars",
      title          = "{\textsc{Collier}: A fortran-based complex one-loop library in
                        extended regularizations}",
      journal        = "Comput. Phys. Commun.",
      volume         = "212",
      year           = "2017",
      pages          = "220-238",
      doi            = "10.1016/j.cpc.2016.10.013",
      eprint         = "1604.06792",
      archivePrefix  = "arXiv",
      primaryClass   = "hep-ph",
      reportNumber   = "FR-PHENO-2016-003, ICCUB-16-016",
      SLACcitation   = "%%CITATION = ARXIV:1604.06792;%%"
}

@article{Anastasiou:2016cez,
      author         = "Anastasiou, Charalampos and Duhr, Claude and Dulat, Falko
                        and Furlan, Elisabetta and Gehrmann, Thomas and Herzog,
                        Franz and Lazopoulos, Achilleas and Mistlberger, Bernhard",
      title          = "{High precision determination of the gluon fusion Higgs
                        boson cross-section at the LHC}",
      journal        = "JHEP",
      volume         = "05",
      year           = "2016",
      pages          = "058",
      doi            = "10.1007/JHEP05(2016)058",
      eprint         = "1602.00695",
      archivePrefix  = "arXiv",
      primaryClass   = "hep-ph",
      reportNumber   = "CP3-16-01, ZU-TH-27-15, NIKHEF-2016-004,
                        CERN-TH-2016-006",
      SLACcitation   = "%%CITATION = ARXIV:1602.00695;%%"
}

@article{Agostinelli:2002hh,
      author         = "Agostinelli, S. and others",
      title          = "{\textsc{Geant4} -- a simulation toolkit}",
      journal        = "Nucl. Instrum. Meth. A",
      volume         = "506",
      year           = "2003",
      pages          = "250",
      doi            = "10.1016/S0168-9002(03)01368-8",
      reportNumber   = "SLAC-PUB-9350, FERMILAB-PUB-03-339",
      SLACcitation   = "%%CITATION = NUIMA,A506,250;%%"
}

@Article{Read:2002hq,
    Author = {Read, Alexander L.},
    Journal = {J. Phys. G},
    Pages = {2693},
    doi = "10.1088/0954-3899/28/10/313",
    Title = "{Presentation of search results: the \(CL_S\) technique}",
    Volume = {28},
    Year = {2002}
}

@Article{Cowan:2010js,
   Author = {Cowan, Glen and Cranmer, Kyle and Gross, Eilam and Vitells, Ofer},
   Title = {{Asymptotic formulae for likelihood-based tests of new physics}},
   Journal = {Eur. Phys. J. C},
   Volume = {71},
   Year = {2011},
   Pages = {1554},
   doi = "10.1140/epjc/s10052-011-1554-0",
   Eprint = "1007.1727",
   Archiveprefix = {arXiv},
   Primaryclass = {physics.data-an},
   related = "Cowan:2010js-err",
   relatedstring  = "Erratum:",
}

@article{Zyla:2020zbs,
    author = "{Particle Data Group} and Zyla, P.A. and others",
    title = "{Review of Particle Physics}",
    journal = "Prog. Theor. Exp. Phys.",
    volume = "2020",
    number = "8",
    pages = "083C01",
    year = "2020",
    doi = "10.1093/ptep/ptaa104",
}


%% file: ATLAS.bib
@Article{PERF-2007-01,
    author         = "{ATLAS Collaboration}",
    title          = "{The ATLAS Experiment at the CERN Large Hadron Collider}",
    journal        = "JINST",
    volume         = "3",
    year           = "2008",
    pages          = "S08003",
    doi            = "10.1088/1748-0221/3/08/S08003",
    primaryClass   = "hep-ex",
}

@Article{SOFT-2010-01,
    author         = "{ATLAS Collaboration}",
    title          = "{The ATLAS Simulation Infrastructure}",
    journal        = "Eur. Phys. J. C",
    volume         = "70",
    year           = "2010",
    pages          = "823",
    doi            = "10.1140/epjc/s10052-010-1429-9",
    eprint         = "1005.4568",
    archivePrefix  = "arXiv",
    primaryClass   = "physics.ins-det",
}

@Article{EXOT-2011-01,
    author         = "{ATLAS Collaboration}",
    title          = "{Search for WH production with a light Higgs boson decaying to prompt electron-jets in proton--proton collisions at \(\sqrt{s} = 7\,\text{TeV}\) with the ATLAS detector}",
    journal        = "New J. Phys.",
    volume         = "15",
    year           = "2013",
    pages          = "043009",
    doi            = "10.1088/1367-2630/15/4/043009",
    reportNumber   = "CERN-PH-EP-2012-365",
    eprint         = "1302.4403",
    archivePrefix  = "arXiv",
    primaryClass   = "hep-ex",
}

@Article{EXOT-2012-16,
    author         = "{ATLAS Collaboration}",
    title          = "{A search for prompt lepton-jets in \(pp\) collisions at \(\sqrt{s} = 7\,\text{TeV}\) with the ATLAS detector}",
    journal        = "Phys. Lett. B",
    volume         = "719",
    year           = "2013",
    pages          = "299",
    doi            = "10.1016/j.physletb.2013.01.034",
    reportNumber   = "CERN-PH-EP-2012-319",
    eprint         = "1212.5409",
    archivePrefix  = "arXiv",
    primaryClass   = "hep-ex",
}

@Article{EXOT-2014-09,
    author         = "{ATLAS Collaboration}",
    title          = "{A search for prompt lepton-jets in \(pp\) collisions at \(\sqrt{s} = 8\,\text{TeV}\) with the ATLAS detector}",
    journal        = "JHEP",
    volume         = "02",
    year           = "2016",
    pages          = "062",
    doi            = "10.1007/JHEP02(2016)062",
    reportNumber   = "CERN-PH-EP-2015-242",
    eprint         = "1511.05542",
    archivePrefix  = "arXiv",
    primaryClass   = "hep-ex",
}

@Article{PERF-2014-03,
    author         = "{ATLAS Collaboration}",
    title          = "{Performance of pile-up mitigation techniques for jets in \(pp\) collisions at \(\sqrt{s} = 8\,\text{TeV}\) using the ATLAS detector}",
    journal        = "Eur. Phys. J. C",
    volume         = "76",
    year           = "2016",
    pages          = "581",
    doi            = "10.1140/epjc/s10052-016-4395-z",
    reportNumber   = "CERN-PH-EP-2015-206",
    eprint         = "1510.03823",
    archivePrefix  = "arXiv",
    primaryClass   = "hep-ex",
}

@Article{STDM-2015-05,
    author         = "{ATLAS Collaboration}",
    title          = "{Measurement of the Inelastic Proton--Proton Cross Section at \(\sqrt{s} = 13\,\text{TeV}\) with the ATLAS Detector at the LHC}",
    journal        = "Phys. Rev. Lett.",
    volume         = "117",
    year           = "2016",
    pages          = "182002",
    doi            = "10.1103/PhysRevLett.117.182002",
    reportNumber   = "CERN-EP-2016-140",
    eprint         = "1606.02625",
    archivePrefix  = "arXiv",
    primaryClass   = "hep-ex",
}

@Article{PERF-2016-04,
    author         = "{ATLAS Collaboration}",
    title          = "{Jet energy scale measurements and their systematic uncertainties in proton--proton collisions at \(\sqrt{s} = 13\,\text{TeV}\) with the ATLAS detector}",
    journal        = "Phys. Rev. D",
    volume         = "96",
    year           = "2017",
    pages          = "072002",
    doi            = "10.1103/PhysRevD.96.072002",
    reportNumber   = "CERN-EP-2017-038",
    eprint         = "1703.09665",
    archivePrefix  = "arXiv",
    primaryClass   = "hep-ex",
}

@Article{TRIG-2016-01,
    author         = "{ATLAS Collaboration}",
    title          = "{Performance of the ATLAS trigger system in 2015}",
    journal        = "Eur. Phys. J. C",
    volume         = "77",
    year           = "2017",
    pages          = "317",
    doi            = "10.1140/epjc/s10052-017-4852-3",
    reportNumber   = "CERN-EP-2016-241",
    eprint         = "1611.09661",
    archivePrefix  = "arXiv",
    primaryClass   = "hep-ex",
}

@Article{EXOT-2017-03,
    author         = "{ATLAS Collaboration}",
    title          = "{Search for long-lived particles in final states with displaced dimuon vertices in \(pp\) collisions at \(\sqrt{s} = 13\,\text{TeV}\) with the ATLAS detector}",
    journal        = "Phys. Rev. D",
    volume         = "99",
    year           = "2019",
    pages          = "012001",
    doi            = "10.1103/PhysRevD.99.012001",
    reportNumber   = "CERN-EP-2018-173",
    eprint         = "1808.03057",
    archivePrefix  = "arXiv",
    primaryClass   = "hep-ex",
}

@Article{DAPR-2018-01,
    author         = "{ATLAS Collaboration}",
    title          = "{ATLAS data quality operations and performance for 2015--2018 data-taking}",
    journal        = "JINST",
    volume         = "15",
    year           = "2020",
    pages          = "P04003",
    doi            = "10.1088/1748-0221/15/04/P04003",
    reportNumber   = "CERN-EP-2019-207",
    eprint         = "1911.04632",
    archivePrefix  = "arXiv",
    primaryClass   = "physics.ins-det",
}

@Article{EGAM-2018-01,
    author         = "{ATLAS Collaboration}",
    title          = "{Electron and photon performance measurements with the ATLAS detector using the 2015--2017 LHC proton--proton collision data}",
    journal        = "JINST",
    volume         = "14",
    year           = "2019",
    pages          = "P12006",
    doi            = "10.1088/1748-0221/14/12/P12006",
    reportNumber   = "CERN-EP-2019-145",
    eprint         = "1908.00005",
    archivePrefix  = "arXiv",
    primaryClass   = "hep-ex",
}

@Article{EXOT-2018-55,
    author         = "{ATLAS Collaboration}",
    title          = "{Search for light neutral particles decaying promptly into collimated pairs of electrons or muons in \(pp\) collisions at \(\sqrt{s} = 13\,\text{TeV}\) with the ATLAS detector}",
    year           = "2024",
    reportNumber   = "CERN-EP-2024-183",
    eprint         = "2407.09168",
    archivePrefix  = "arXiv",
    primaryClass   = "hep-ex",
}

@Article{HDBS-2018-55,
    author         = "{ATLAS Collaboration}",
    title          = "{Search for Higgs bosons decaying into new spin-0 or spin-1 particles in four-lepton final states with the ATLAS detector with \(139\,\text{fb}^{-1}\) of \(pp\) collision data at \(\sqrt{s} = 13\,\text{TeV}\)}",
    journal        = "JHEP",
    volume         = "03",
    year           = "2022",
    pages          = "041",
    doi            = "10.1007/JHEP03(2022)041",
    reportNumber   = "CERN-EP-2021-193",
    eprint         = "2110.13673",
    archivePrefix  = "arXiv",
    primaryClass   = "hep-ex",
}

@Article{MUON-2018-03,
    author         = "{ATLAS Collaboration}",
    title          = "{Muon reconstruction and identification efficiency in ATLAS using the full Run~2 \(pp\) collision data set at \(\sqrt{s} = 13\,\text{TeV}\)}",
    journal        = "Eur. Phys. J. C",
    volume         = "81",
    year           = "2021",
    pages          = "578",
    doi            = "10.1140/epjc/s10052-021-09233-2",
    reportNumber   = "CERN-EP-2020-199",
    eprint         = "2012.00578",
    archivePrefix  = "arXiv",
    primaryClass   = "hep-ex",
}

@Article{PIX-2018-001,
    author         = "Abbott, B. and others",
    title          = "{Production and integration of the ATLAS Insertable B-Layer}",
    journal        = "JINST",
    volume         = "13",
    year           = "2018",
    pages          = "T05008",
    doi            = "10.1088/1748-0221/13/05/T05008",
    eprint         = "1803.00844",
    archivePrefix  = "arXiv",
    primaryClass   = "physics.ins-det",
}

@Article{TRIG-2018-01,
    author         = "{ATLAS Collaboration}",
    title          = "{Performance of the ATLAS muon triggers in Run~2}",
    journal        = "JINST",
    volume         = "15",
    year           = "2020",
    pages          = "P09015",
    doi            = "10.1088/1748-0221/15/09/p09015",
    reportNumber   = "CERN-EP-2020-031",
    eprint         = "2004.13447",
    archivePrefix  = "arXiv",
    primaryClass   = "physics.ins-det",
}

@Article{TRIG-2018-05,
    author         = "{ATLAS Collaboration}",
    title          = "{Performance of electron and photon triggers in ATLAS during LHC Run~2}",
    journal        = "Eur. Phys. J. C",
    volume         = "80",
    year           = "2020",
    pages          = "47",
    doi            = "10.1140/epjc/s10052-019-7500-2",
    reportNumber   = "CERN-EP-2019-169",
    eprint         = "1909.00761",
    archivePrefix  = "arXiv",
    primaryClass   = "hep-ex",
}

@Article{EXOT-2019-05,
    author         = "{ATLAS Collaboration}",
    title          = "{Search for light long-lived neutral particles that decay to collimated pairs of leptons or light hadrons in \(pp\) collisions at \(\sqrt{s} = 13\,\text{TeV}\) with the ATLAS detector}",
    journal        = "JHEP",
    volume         = "06",
    year           = "2023",
    pages          = "153",
    doi            = "10.1007/JHEP06(2023)153",
    reportNumber   = "CERN-EP-2022-054",
    eprint         = "2206.12181",
    archivePrefix  = "arXiv",
    primaryClass   = "hep-ex",
}

@Article{HIGG-2019-23,
    author         = "{ATLAS Collaboration}",
    title          = "{Search for boosted diphoton resonances in the \(10\) to \(70\,\text{GeV}\) mass range using \(138\,\text{fb}^{-1}\) of \(13\,\text{TeV}\) \(pp\) collisions with the ATLAS detector}",
    journal        = "JHEP",
    volume         = "07",
    year           = "2023",
    pages          = "155",
    doi            = "10.1007/JHEP07(2023)155",
    reportNumber   = "CERN-EP-2022-183",
    eprint         = "2211.04172",
    archivePrefix  = "arXiv",
    primaryClass   = "hep-ex",
}

@Article{TRIG-2019-03,
    author         = "{ATLAS Collaboration}",
    title          = "{The ATLAS inner detector trigger performance in \(pp\) collisions at \(13\,\text{TeV}\) during LHC Run~2}",
    journal        = "Eur. Phys. J. C",
    volume         = "82",
    year           = "2022",
    pages          = "206",
    doi            = "10.1140/epjc/s10052-021-09920-0",
    reportNumber   = "CERN-EP-2021-076",
    eprint         = "2107.02485",
    archivePrefix  = "arXiv",
    primaryClass   = "hep-ex",
}

@Article{DAPR-2021-01,
    author         = "{ATLAS Collaboration}",
    title          = "{Luminosity determination in \(pp\) collisions at \(\sqrt{s} = 13\,\text{TeV}\) using the ATLAS detector at the LHC}",
    journal        = "Eur. Phys. J. C",
    volume         = "83",
    year           = "2023",
    pages          = "982",
    doi            = "10.1140/epjc/s10052-023-11747-w",
    reportNumber   = "CERN-EP-2022-281",
    eprint         = "2212.09379",
    archivePrefix  = "arXiv",
    primaryClass   = "hep-ex",
}

@Article{HIGG-2021-23,
    author         = "{ATLAS Collaboration}",
    title          = "{A detailed map of Higgs boson interactions by the ATLAS experiment ten years after the discovery}",
    journal        = "Nature",
    volume         = "607",
    year           = "2022",
    pages          = "52--59",
    doi            = "10.1038/s41586-022-04893-w",
    reportNumber   = "CERN-EP-2022-057",
    eprint         = "2207.00092",
    archivePrefix  = "arXiv",
    primaryClass   = "hep-ex",
    related        = "HIGG-2021-23-err",
    relatedstring  = "Erratum:",
}

@Article{EXOT-2022-15,
    author         = "{ATLAS Collaboration}",
    title          = "{Search for light long-lived neutral particles from Higgs boson decays via vector-boson-fusion production from \(pp\) collisions at \(\sqrt{s} = 13\,\text{TeV}\) with the ATLAS detector}",
    journal        = "Eur. Phys. J. C",
    volume         = "84",
    year           = "2023",
    pages          = "719",
    doi            = "10.1140/epjc/s10052-024-12902-7",
    reportNumber   = "CERN-EP-2023-226",
    eprint         = "2311.18298",
    archivePrefix  = "arXiv",
    primaryClass   = "hep-ex",
}

@Article{SOFT-2022-02,
    author         = "{ATLAS Collaboration}",
    title          = "{Software and computing for Run~3 of the ATLAS experiment at the LHC}",
    year           = "2024",
    reportNumber   = "CERN-EP-2024-100",
    eprint         = "2404.06335",
    archivePrefix  = "arXiv",
    primaryClass   = "hep-ex",
}

@Booklet{ATL-SOFT-PUB-2023-001,
    author         = "{ATLAS Collaboration}",
    title          = "{ATLAS Computing Acknowledgements}",
    howpublished   = "{ATL-SOFT-PUB-2023-001}",
    url            = "https://cds.cern.ch/record/2869272",
    year           = "2023",
}

@Report{ATLAS-TDR-19,
    author         = "{ATLAS Collaboration}",
    title          = "{ATLAS Insertable B-Layer: Technical Design Report}",
    type           = "ATLAS-TDR-19; CERN-LHCC-2010-013",
    year           = "2010",
    url            = "https://cds.cern.ch/record/1291633",
    related        = "ATLAS-TDR-19-addm",
    relatedstring  = "Addendum:",
}

@Article{LUCID2,
    author         = "Avoni, G. and others",
    title          = "{The new LUCID-2 detector for luminosity measurement and monitoring in ATLAS}",
    journal        = "JINST",
    volume         = "13",
    year           = "2018",
    pages          = "P07017",
    doi            = "10.1088/1748-0221/13/07/P07017",
}


%% file: CMS.bib
@Article{CMS-EXO-11-013,
    author         = "{CMS Collaboration}",
    title          = "{Search for light resonances decaying into pairs of muons as a signal of new physics}",
    journal        = "JHEP",
    volume         = "07",
    year           = "2011",
    pages          = "098",
    doi            = "10.1007/JHEP07(2011)098",
    reportNumber   = "CERN-PH-EP-2011-064",
    eprint         = "1106.2375",
    archivePrefix  = "arXiv",
    primaryClass   = "hep-ex",
}

@Article{CMS-EXO-12-012,
    author         = "{CMS Collaboration}",
    title          = "{Search for a non-standard-model Higgs boson decaying to a pair of new light bosons in four-muon final states}",
    journal        = "Phys. Lett. B",
    volume         = "726",
    year           = "2013",
    pages          = "564",
    doi            = "10.1016/j.physletb.2013.09.009",
    reportNumber   = "CERN-PH-EP-2012-292",
    eprint         = "1210.7619",
    archivePrefix  = "arXiv",
    primaryClass   = "hep-ex",
}

@Article{CMS-HIG-13-010,
    author         = "{CMS Collaboration}",
    title          = "{A search for pair production of new light bosons decaying into muons}",
    journal        = "Phys. Lett. B",
    volume         = "752",
    year           = "2016",
    pages          = "146",
    doi            = "10.1016/j.physletb.2015.10.067",
    reportNumber   = "CERN-PH-EP-2015-116",
    eprint         = "1506.00424",
    archivePrefix  = "arXiv",
    primaryClass   = "hep-ex",
}

@Article{CMS-HIG-19-007,
    author         = "{CMS Collaboration}",
    title          = "{Search for low-mass dilepton resonances in Higgs boson decays to four-lepton final states in proton--proton collisions at \(\sqrt{s} = 13\,\text{TeV}\)}",
    journal        = "Eur. Phys. J. C",
    volume         = "82",
    year           = "2022",
    pages          = "290",
    doi            = "10.1140/epjc/s10052-022-10127-0",
    reportNumber   = "CERN-EP-2021-093",
    eprint         = "2111.01299",
    archivePrefix  = "arXiv",
    primaryClass   = "hep-ex",
}

@Article{CMS-EXO-20-014,
    author         = "{CMS Collaboration}",
    title          = "{Search for long-lived particles decaying into muon pairs in proton--proton collisions at \(\sqrt{s}\ = 13\,\text{TeV}\) collected with a dedicated high-rate data stream}",
    journal        = "JHEP",
    volume         = "04",
    year           = "2022",
    pages          = "062",
    doi            = "10.1007/JHEP04(2022)062",
    reportNumber   = "CERN-EP-2021-266",
    eprint         = "2112.13769",
    archivePrefix  = "arXiv",
    primaryClass   = "hep-ex",
}

@Article{CMS-EXO-21-005,
    author         = "{CMS Collaboration}",
    title          = "{Search for direct production of GeV-scale resonances decaying to a pair of muons in proton--proton collisions at \(\sqrt{s} = 13\,\text{TeV}\)}",
    journal        = "JHEP",
    volume         = "12",
    year           = "2023",
    pages          = "070",
    doi            = "10.1007/JHEP12(2023)070",
    reportNumber   = "CERN-EP-2023-165",
    eprint         = "2309.16003",
    archivePrefix  = "arXiv",
    primaryClass   = "hep-ex",
}

@Article{CMS-HIG-22-001,
    author         = "{CMS Collaboration}",
    title          = "{A portrait of the Higgs boson by the CMS experiment ten years after the discovery}",
    journal        = "Nature",
    volume         = "607",
    year           = "2022",
    pages          = "60--68",
    doi            = "10.1038/s41586-022-04892-x",
    reportNumber   = "CERN-EP-2022-039",
    eprint         = "2207.00043",
    archivePrefix  = "arXiv",
    primaryClass   = "hep-ex",
    related        = "CMS-HIG-22-001-err",
    relatedstring  = "Erratum:",
}


%% file: ConfNotes.bib
@Booklet{ATLAS-CONF-2015-029,
    author         = "{ATLAS Collaboration}",
    title          = "{Selection of jets produced in \(13~\text{TeV}\) proton--proton collisions with the ATLAS detector}",
    howpublished   = "{ATLAS-CONF-2015-029}",
    url            = "https://cds.cern.ch/record/2037702",
    year           = "2015",
}


%% file: PubNotes.bib
@Booklet{ATL-PHYS-PUB-2014-021,
    author         = "{ATLAS Collaboration}",
    title          = "{ATLAS Pythia~8 tunes to \(7~\text{TeV}\) data}",
    howpublished   = "{ATL-PHYS-PUB-2014-021}",
    url            = "https://cds.cern.ch/record/1966419",
    year           = "2014",
}

@Booklet{ATL-PHYS-PUB-2015-026,
    author         = "{ATLAS Collaboration}",
    title          = "{Vertex Reconstruction Performance of the ATLAS Detector at \(\sqrt{s} = 13~\text{TeV}\)}",
    howpublished   = "{ATL-PHYS-PUB-2015-026}",
    url            = "https://cds.cern.ch/record/2037717",
    year           = "2015",
}

@Booklet{ATL-PHYS-PUB-2015-036,
    author         = "{ATLAS Collaboration}",
    title          = "{Properties of jets and inputs to jet reconstruction and calibration with the ATLAS detector using proton--proton collisions at \(\sqrt{s} = 13~\text{TeV}\)}",
    howpublished   = "{ATL-PHYS-PUB-2015-036}",
    url            = "https://cds.cern.ch/record/2044564",
    year           = "2015",
}

@Booklet{ATL-PHYS-PUB-2016-017,
    author         = "{ATLAS Collaboration}",
    title          = "{The Pythia~8 A3 tune description of ATLAS minimum bias and inelastic measurements incorporating the Donnachie--Landshoff diffractive model}",
    howpublished   = "{ATL-PHYS-PUB-2016-017}",
    url            = "https://cds.cern.ch/record/2206965",
    year           = "2016",
}

@Booklet{ATL-PHYS-PUB-2016-020,
    author         = "{ATLAS Collaboration}",
    title          = "{Studies on top-quark Monte Carlo modelling for Top2016}",
    howpublished   = "{ATL-PHYS-PUB-2016-020}",
    url            = "https://cds.cern.ch/record/2216168",
    year           = "2016",
}
